\documentclass[12pt,reqno]{amsart}
\numberwithin{equation}{section}
\usepackage{amsmath}
\usepackage{amssymb,amsthm}
\usepackage{amsbsy,amsfonts,amscd}
\usepackage{epsfig}
\usepackage{psfrag}
\usepackage{color,verbatim}
\usepackage{graphics,graphicx}
\usepackage{ulem}

\usepackage{fullpage}

\usepackage{xcolor}
\usepackage{pstricks,pst-node}
\usepackage{tikz}
\usetikzlibrary{backgrounds}
\usetikzlibrary{patterns,fadings}
\usetikzlibrary{arrows,decorations.pathmorphing}
\usetikzlibrary{shapes}
\usetikzlibrary{calc}

\usepackage{enumerate}

\definecolor{brightcerulean}{rgb}{0.11, 0.67, 0.84}
\definecolor{cerulean}{rgb}{0.0, 0.48, 0.65}
\definecolor{Gray}{rgb}{0.5, 0.5, 0.5}

\newcommand\bx{{\mathbf x}}
\newcommand\by{{\mathbf y}}

\newcommand\bk{{\mathbf k}}

\newcommand\bn{{\mathbf n}}

\newcommand\RR{{\mathbb R}}

\newcommand\ZZ{{\mathbb Z}}

\renewcommand\ll{\langle\!\langle}
\renewcommand\gg{\rangle\!\rangle}

\renewcommand\div{{\rm{div}}\;}

\newcommand{\mc}[1]{{\mathcal #1}}

\newcommand{\bb}[1]{{\mathbb #1}}

\begin{document}

\title[]{Density large deviations for multidimensional stochastic hyperbolic conservation laws}

\author{J. Barr\'e}

\address{MAPMO - UMR CNRS 7349, F\'ed\'eration Denis Poisson\\
Universit\'e d'Orl\'eans, Collegium Sciences et Techniques \\
B\^atiment de math\'ematiques - Route de Chartres\\
B.P. 6759 - 45067 Orl\'eans cedex 2 
FRANCE\\
et Institut Universitaire de France}
\email{{\tt julien.barre@univ-orleans.fr}}

\author{C.Bernardin}
\address{Universit\'e C\^ote d'Azur, CNRS, LJAD\\
Parc Valrose\\
06108 NICE Cedex 02, France}
\email{{\tt cbernard@unice.fr}}

\author{R. Chetrite}

\address{Universit\'e C\^ote d'Azur, CNRS, LJAD\\
Parc Valrose\\
06108 NICE Cedex 02, France}
\email{{\tt raphael.chetrite@unice.fr}}

\thanks{}

\date{\today.}
\begin{abstract}
We investigate the density large deviation function for a multidimensional conservation law in the vanishing viscosity limit, when the probability concentrates on weak solutions of a hyperbolic conservation law. When the mobility and diffusivity matrices are proportional, i.e. an Einstein-like relation is satisfied, the problem has been solved in \cite{BM}. When this proportionality does not hold, we compute explicitly the large deviation function for a step-like density profile, and we show that the associated optimal current has a non trivial structure. We also derive a lower bound for the large deviation function, valid for a general weak solution, and leave the general large deviation function upper bound as a conjecture.
\end{abstract}

\keywords{Large Deviations Principle, Stochastic conservation laws, Kawasaki dynamics, Active particles.} 


\thanks{ }
\newtheorem{theorem}{Theorem}[section]
\newtheorem{proposition}[theorem]{Proposition}
\newtheorem{lemma}[theorem]{Lemma}
\newtheorem{corollary}[theorem]{Corollary}
\newtheorem{assumption}[theorem]{Assumption}
\theoremstyle{remark}
\newtheorem{remark}[theorem]{Remark}
\newtheorem{definition}[theorem]{Definition}
\makeatletter
\newcommand{\cqfd}{{\unskip\kern 6pt\penalty 500
\raise -2pt\hbox{\vrule\vbox to 6pt{\hrule width 6pt
\vfill\hrule}\vrule}\par}}

\maketitle
\section{Introduction}
\label{intro}
%

Systems of interacting particles are often described on large scales by a partial differential equation.
This PDE results in some sense from a Law of Large Number, and it is sometimes important to go beyond this level, and to include a description of the fluctuations around the most probable evolution \cite{S}, expressed by this PDE. Many studies have been devoted to this kind of description for  driven diffusive systems, leading to what is called `Macroscopic Fluctuation Theory", a cornerstone of modern out of equilibrium statistical physics (for a recent review, see \cite{MFT}). By contrast, less is known for systems whose macroscopic description involves a transport equation, which is the main set up here.    
In this article, we approach this second set-up with the first one and for this we consider a generic multi-dimensional transport-diffusion system in the limit of small diffusion: the hydrodynamic for the density{\footnote{Depending on the context $\rho$ has to be interpreted as a density, an energy ... }} $\rho:=\rho(t,\bx) \ge 0$, $\bx \in \RR^d$, with $d\ge 2$, is described by the parabolic equation
\begin{equation}
\label{eq:pde1}
\partial_t \rho + \div j_\nu (\rho) =0, \quad j_\nu (\rho)= - \frac{1}{\nu}D(\rho) \nabla \rho + f (\rho).
\end{equation}
Here $D(\rho)$ is a square symmetric matrix called diffusivity, $f(\rho)$ is a $d$-dimensional vector called hyperbolic flux. The parameter $\nu>0$ regulates the strength of the diffusion and will tend to infinity later on: in this limit, \eqref{eq:pde1} becomes a scalar hyperbolic conservation law. The solution of (\ref{eq:pde1}) will be denoted by $\rho^\nu$.

To take into account the fluctuations around this typical behavior we have to replace the previous PDE by the SPDE \cite{LL, HH, S}
\begin{equation}
\label{eq:spde1}
\partial_t \rho + \div {J}_\nu (\rho) =0, \quad J_\nu (\rho)= - \frac{1}{\nu}D(\rho) \nabla \rho + f (\rho) + \sqrt{\cfrac{\sigma (\rho)}{N\nu}} \, \eta.
\end{equation}
Here $\eta$ is a  space-time Gaussian white noise, $\sigma (\rho)$ is a symmetric matrix valued function called mobility and $N$ is the number of particles in the interacting particle system which tends to infinity. Examples of interacting particle systems, among many others, which are described by stochastic conservation laws (\ref{eq:spde1}) are the following : 
\begin{itemize}
\item {\bf{Driven Kawasaki exchange dynamics}}: These dynamics \cite{Ka} are jump Markov processes $(\eta_t)_{t\ge 0}$ with state space $\Omega_{\Lambda}=\{0,1\}^\Lambda$, $\Lambda$ a sub-lattice of $\ZZ^d$. For a configuration $\eta:=\{ \eta(x) \, ; \, x \in \Lambda\}$ we interpret $\eta(x)=1$ as the presence of a particle at site $x$ and $\eta(x)=0$ as its absence. The dynamics are such that the number of particles is locally conserved. They were first introduced as {\textit{reversible}} dynamics w.r.t. the Gibbs measure with some Hamiltonian $H$ by imposing the corresponding detailed balance condition on the rates $c_0 (x,y;\eta)$, $x,y \in \Lambda$, $\eta \in \Omega_{\Lambda}$. Yet, adding some external constant and homogeneous electric field $E$ to such a system {\footnote{This means that the initial jump rates $c_0$ are modified into $c_E (x,y,\eta)$ such that the ``local detailed balance" $ c_E (x,y,\eta) =c_E(x,y,\eta^{xy})  e^{-(H (\eta^{xy}) -H(\eta)) -E \dot (x-y) (\eta (x) -\eta (y))}$, with $\eta^{xy}$ the configuration obtained by exchanging $\eta(x)$ with $\eta(y)$, is satisfied. }}  results in a nonequilibrium stationary state with a non-zero average flux of particles, which moreover is usually not Gibbsian. If the rates are anisotropic then the matrices $D$ and $\sigma$ are not proportional and long-range correlations are expected, despite the fact that the dynamics is only local. In the case of isotropic jump rates, the matrices $D$ and $\sigma$ are proportional and the stationary state has short range correlations \cite{GLMS}, \cite{MR},\cite{BFG}. 
\item {\bf{Active particles}}: Different biological systems, from bacteria to flocks of mammals, are described by interacting particles, each one of them being self-propelled.  
The macroscopic description of such systems is in general more complicated than \eqref{eq:pde1}, including several PDEs \cite{Bertin,Degond}; \cite{Chate} provides an example where the finite size noise is kept in the final equations. Nevertheless, simplified models can fit exactly in the framework \eqref{eq:spde1} \cite{nous}. For these systems, there is no reason that noise and diffusion satisfy an Einstein relation, and $\sigma$ and $D$ are not proportional in general.\\
\end{itemize}

Due to the nonlinearity,  the SPDE \eqref{eq:spde1}  is in general ill defined and need to be properly renormalized \cite{DeD, FJL,H}. Here, a precise meaning is given by restricting to the small noise limit, and by interpreting \eqref{eq:spde1} in the large deviation framework \cite{OM, St, FW, ZJ, MFT}. Let us fix a horizon time $T>0$ and define $\Omega=[0,T] \times \RR^d$.  The couple $(j,\rho)$ satisfies then a large deviation principle with speed $N\nu$ on the time window $[0,T]$. Moreover the Large Deviations Function (LDF) can be obtained formally as  \cite{OM, St, FW, ZJ, MFT} 
 \begin{equation}
 \begin{split}
 & {\mc I}^\nu_{[0,T]} (j, \rho)= {\cfrac{1}{2}} \int_\Omega \left \langle \big[ j - j_\nu (\rho) \big] \, , \,  \sigma^{-1} (\rho) \big[ j -j_\nu (\rho) \big] \right \rangle \; dt d\bx 
  \label{eq:ldp}
 \end{split}
 \end{equation}
 if the constraint
 \begin{equation}
 \partial_t \rho + \div j =0 
  \label{eq:c}
 \end{equation}
is satisfied and equal to infinity otherwise. Hereafter $\langle \cdot, \cdot \rangle$ denotes the usual scalar product of $\RR^d$. This LDF describes the cost to observe during a time window $[0,T]$ a density profile $\rho$ and a current profile $j$ for the underlying microscopic system. 
 

We are interested{\footnote{A natural problem to investigate is also the LDF for the current. See \cite{PGH, TGH,TPGH} for studies in this direction.}} in the density LDF $H^{\nu}_{[0,T]}$ which describes the cost to observe an atypical density profile over a time interval $[0,T]$; it is  related to ${\mc I}^\nu$ by a contraction principle \cite{V2, dH,V2}, over the admissible currents $j$:
\begin{equation}
\label{eq:Hnunu}
H^{\nu}_{[0,T]} (\rho) = \inf_{j~s.t.~\partial_t \rho +\div j=0} {\mc I}_{[0,T]}^{\nu} (j,\rho).
\end{equation}
We will focus on the limiting form of $H^{\nu}_{[0,T]}$ as $\nu \to \infty$, i.e. for systems whose typical behavior is a scalar hyperbolic conservation law. Taking formally the limit $\nu \to \infty$, it is easy to convince oneself that the density LDF vanishes
for any $\rho$ such that $\partial_t \rho +\div f(\rho)=0$. Indeed, for such a $\rho$, the choice $j=f(\rho)$ fulfills the constraint equation  \eqref{eq:c} in the $\nu \to \infty$ formal limit of the equation \eqref{eq:pde1}, and the integral appearing in the definition of $H^{\nu}$ vanishes in the $\nu \to \infty$ limit. In other words, the probability concentrates on all weak solutions of the hyperbolic conservation law \cite{M}. There are many such weak solutions, and from the point of view of fluctuations, this limit of the LDF misses some interesting physical properties. To go further we consider the limit of the scaled LDF: we
look for a large deviation principle with speed $N$, and define  
\[
{\bb H}^{\infty}_{[0,T]} (\rho) =\lim_{\nu \to \infty} \nu H^{\nu}_{[0,T]} (\rho).
\]
Clearly, ${\bb H}^{\infty}$ is infinite if $\rho$ is not a weak solution of the conservation law. Our goal is to compute ${\bb H}^{\infty}$ for $\rho$ such a weak solution. In full rigor, the above limit shall be understood in the sense of $\Gamma$-convergence{\footnote{\label{foot:gamma} A sequence of functional $F^\nu: \chi \to \RR$ defined on some topological space $\chi$ $\Gamma$-converges to $F:\chi \to\RR$ if 1) for any $x \in \chi$ and any sequence $x^\nu \to x$, $\liminf F^\nu (x^\nu) \ge F (x)$ ($\Gamma$-liminf inequality) and 2) there exists a sequence $x^\nu \to x$ such that $\limsup F^\nu (x^\nu) \le F(x)$ ($\Gamma$-limsup inequality).}} which is the right notion to deal with convergence of variational problems \cite{DaM}. 

As far as we know, this problem has been investigated mainly in the one dimensional case (\cite{BD} considers a specific example with $\sigma=f$ concave; \cite{BBMN} treats the general case). The $d$-dimensional case is solved in \cite{BM} under the restrictive hypothesis that the matrices $D$ and $\sigma$ are proportional: we will see that far from being a technical condition, this is a fundamental hypothesis. Motivated in particular by active particles systems, the aim of this paper is to make progresses in the generic $d$-dimensional case.

As explained above, we expect ${\bb H}^{\infty}_{[0,T]} (\rho)$ to be infinite if $\rho$ is not a weak solution (see Section \ref{subsec:weak}) of the scalar conservation law $\partial_t \rho + \div f(\rho) =0.$ Hence we take $\rho$ to be such a weak solution.
Weak solutions are usually continuous functions apart from some codimension 1 time-space manifold $J_\rho \subset \Omega$. A point $(t,\bx) \in J_{\rho}$ is classified as a shock or an anti-shock according to the fact that it dissipates or produces entropy {\footnote{Notice that the notion of entropy production discussed here is a PDE concept different from the corresponding physical concept.}} , see Section \ref{sec:weak_entrop}. In the one dimensional case, it  has been proved \cite{J,V,M} that only the anti-shocks give a contribution to ${\bb H}^{\infty}$, and that the entropy production has to be measured using the inverse of the susceptibility  $D/\sigma$ . 
Furthermore, these contributions add up, i.e. must be integrated along $J_\rho$, and the infinitesimal contribution at $(t,\bx)$ can be obtained by approximating the weak solution $\rho$ by a moving step propagating in the normal direction to $J_\rho$ at $(t,\bx)$. 

Note that, since in the $d\ge 2$ dimensional case we are dealing now with matrices, $D/\sigma$ in general makes no sense, so that we have to introduce a new measure of entropy production. Our main result is to derive a formula (\ref{eq:JVgen_movingstep}) giving ${\bb H}^{\infty}_{[0,T]} (\rho)$ when the weak solution $\rho$ is a moving step. If the additivity principle proved in the one-dimensional case also holds in the multi-dimensional case, which is not obvious, we can deduce a formula (\ref{eq:conj}) for  ${\bb H}^{\infty}_{[0,T]}$.

The article is organized as follows: in Section~\ref{sec:weak_entrop}, we gather the definitions and results on hyperbolic scalar conservation laws which will be needed later on; although this material is by no means original, it is not necessarily well-known to statistical physicists. Then, in Section~\ref{sec:densityLDF}, we derive our main results: i) a (quite formal) lower bound for the probability of observing a generic weak solution in terms of the entropy production in Section~\ref{sec:lowerbounds};  ii) a lower and upper bound for the probability to observe a given moving step function, which coincides in this particular case with the previous lower bound, in Section~\ref{sec:conjecture_checking_step}; iii) the most probable current associated to this moving step function in  Section~\ref{UB_entrop_split}: 
contrary to the case when an Einstein relation holds, it has a non trivial structure. The upper bound for a generic weak solution, which, together with point i), would give the probability to observe a generic weak solution, is left as a conjecture: it depends on the validity of an additive principle, which we state explicitly in Section~\ref{sec:additive_principle}. We emphasize the differences with the case where an Einstein relation holds. Some technical points are detailed in the appendices.

Finally, we note that there has been a number of studies on ``stochastic scalar conservation laws" (see for instance \cite{Kim,Feng,Debussche}); in these references, the noise term is non conservative, and the questions addressed are quite different.  \\

 In the rest of the paper we only consider the $d=2$ case for simplicity but our results extend in the multidimensional case.

\section{Solutions of hyperbolic conservation laws: Formulation, Lyapounov functions and Entropy production} 
\label{sec:weak_entrop}

The aim of this section is to present the mathematical concepts of weak and entropic solutions of a scalar conservation law and to explain the relevance of these notions from a physical viewpoint.

\subsection{Generalities and Entropy production}
\label{subsec:weak}

Consider the $2d$-scalar conservation law
\begin{equation}
\label{eq:scl0}
\partial_t \rho + {\rm{div}} \, f(\rho) =0.
\end{equation}
Since this equation does not admit classical (smooth) solutions {\footnote{In general, smooth solutions only exist for a short time and develop shocks. For particular initial conditions, smooth solutions exist but this is quite exceptional.}}, it has to be interpreted in a weak sense, i.e. by integrating w.r.t. smooth test functions $\varphi (t,\bx)$, $(t,\bx) \in \Omega=[0,T] \times \RR^2$, and transporting the partial derivatives on the test function by a formal integration by parts:
\begin{equation}
\label{eq:scl1}
\int_\Omega \; \left\{ \rho\,  \partial_t \varphi + \langle f (\rho),  \nabla \varphi \rangle \right\} \, dt \, d \bx  =0.
\end{equation} 
Therefore we say that a function $\rho (t,\bx)$ is a weak solution of (\ref{eq:scl0}) if it satisfies (\ref{eq:scl1}) for any smooth test function $\varphi$. But since there exist several weak solutions to select the ``physical" one we shall impose an extra condition to weak solution to restore uniqueness. Here ``physical" means that it can be derived by a space-time coarse graining procedure from the underlying microscopic model as the typical macroscopic profile observed in a suitable time scale.

In the PDE's literature, a scalar function $\eta$ on $\RR$ is called ``entropy" for (\ref{eq:scl0}) if  (\ref{eq:scl0}) is compatible with an extra conservation law in the form 
$$\partial_t \eta (\rho) + {\rm{div}} \, {q} (\rho) =0.$$
In the context of systems of conservation laws, entropies are quite difficult to obtain but in the scalar case, any function $\eta$ is an entropy. Indeed, it is sufficient to define ${q}=({q}_x, {q}_y)$, which is a vector-valued function on $\RR$, such that $q' = \eta^{\prime} f'$. The vector $q$ is called the conjugated entropy flux to the entropy $\eta$. 

A weak solution is called entropic if for each entropy-entropy flux pair $(\eta, {q})$ with $\eta$ convex, the inequality 
\begin{equation}
\label{eq:entr-ineq}
\partial_t \eta(\rho) + {\rm{div}}\,   {q} (\rho) \le 0
\end{equation}
holds in the sense of distributions, i.e. for any positive test function $\varphi (t, \bx)$
\begin{equation*}
\int_\Omega \left\{ \partial_t \varphi \, \eta (\rho) + \langle q(\rho) , \nabla \varphi \rangle \right\} dt d\bx \ge 0.
\end{equation*}
Observe that if $\rho$ is a smooth (classical) solution then the previous inequality becomes an equality. Existence and uniqueness of an entropic solution has been proved under generic conditions \cite{Daf,Se}. Thus in this sense the entropic solution is the only weak solution dissipating entropy. 

The relevance of the entropic solution for an asymmetric microscopic system with one conservation law is that among all the weak solutions, this is the solution which describes the {\textit{typical}} behavior of the microscopic system. This has been rigorously proved for only few asymmetric Kawasaki exchange dynamics in dimension $d\ge 2$ such that the Asymmetric Exclusion Process (\cite{Re}). Weak solutions are usually considered as irrelevant but as proved in \cite{J,V,BBMN} they play a special role to understand the fluctuations of the density at the macroscopic level. 

We extend the notion of entropy-entropy flux pair by defining an entropy sampler ${\mc V}(\rho, t, \bx)$ and its conjugated entropy flux sampler ${\mc Q}(\rho, t, \bx)$ as scalar and vector-valued functions such that for any $(t,\bx) \in \Omega$,  $({\mc V} (\cdot, t, \bx), {\mc Q} (\cdot, t, \bx))$ is an entropy-entropy flux pair. This means that for all $(t,\bx) \in \Omega$:
\[
{\mc Q}^\prime \left(\rho\left(t,\bx\right),t,\bx\right)={\mc V}^{\prime}\left(\rho\left(t,\bx\right),t,\bx\right)f^{\prime} \left(\rho\left(t,\bx\right)\right),
\]
where $'$ denotes the derivative with respect to the $\rho$ argument. An example is the factorized case ${\mc V} (\rho,t,\bx) =\eta(\rho) \varphi(t, \bx)$, ${\mc Q}(\rho,t,\bx) = q(\rho) \varphi(t,\bx)$ where
$q' = \eta^{\prime} f'$ and $\varphi$ is an arbitrary smooth function. Then the {\textit{${\mc V}$-sampled entropy production}} {\footnote{Observe that the notion of entropy production introduced here is a PDE concept which is quite different from the corresponding ``physical" concept \cite{KP, DM, Ma, GL}. In particular, following the mathematical convention, entropy typically increases.}} on the time interval $[0,T]$ of a function $\rho(t, \bx)$, weak solution of \eqref{eq:scl0}, is defined as the real number with non-definite sign
\begin{equation}
\label{eq:entprod}
{\mc P}_{\mc V} (\rho) := - \int_{\Omega} \left[ (\partial_t {\mc V} ) \, (\rho(t,\bx), t, \bx) +(\nabla_{\bx} \cdot {\mc Q})\, ( \rho(t, \bx), t, \bx) \right]  \; dt d{\bx},
\end{equation}
where $\nabla_\bx$ the gradient with respect to the third (space) variable of $\mc Q$. ${\mc V}$-sampled entropy production will play an important role in the large deviation function for the density studied in the next sections.

\subsection{Viscosity solution}

The entropic solution of the hyperbolic conservation law \eqref{eq:scl0} is denoted by $\rho^\infty$. It can be obtained as the vanishing viscosity limit $\rho=\lim_{\nu \to \infty} {\hat \rho}^{\nu}$ of the smooth solution ${\hat \rho}^{\nu}$ of (see \cite{Daf,Se})
$$\partial_t {\hat \rho}^{\nu} + {\rm div} f({\hat \rho}^{\nu}) = \tfrac{1}{\nu} {\rm div} \big ( {\hat D}({\hat \rho}^{\nu}) \nabla {\hat \rho}^{\nu})\big)$$
where $\hat D$ is a uniformly elliptic matrix-valued function, which means that ${\hat D}(\rho)\ge \kappa>0$ for any density $\rho$. In particular, under the ellipticity assumption on $D$, the solution $\rho^\nu$ of (\ref{eq:pde1}), converges to the entropic solution $\rho^\infty$.

The vanishing viscosity approach also explains the inequality (\ref{eq:entr-ineq}) since (\ref{eq:entr-ineq}) holds for ${\hat \rho}^\nu$ up to a term vanishing in the $\nu \to \infty$ limit; the inequality will thus persist in this limit.

\subsection{Quasi-potential and Lyapounov function}

For $\nu>0$ (resp. $\nu=\infty$), the quasi-potential $V^{\nu}$ (resp. ${\bb V}^{\infty}$) associated to the dynamical LDF $H^{\nu}$ (resp. $ {\bb H}^{\infty}$) is defined by 
\begin{equation}
\label{eq:qp1}
V^{\, \nu} (\gamma) := \inf_{\rho}  \; H^{\, \nu}_{(-\infty, 0]}\; (\rho) , \quad \Big({\rm resp.} \;  {\bb V}^{\, \infty} (\gamma) := \inf_{\rho}  \; {\bb H}^{\, \infty}_{(-\infty, 0]}\; (\rho) \, \Big)
\end{equation}
where $\gamma:\RR^2 \to \RR$ is a time independent density profile and the infimum is carried on the set of time-space density profiles $\rho:=\rho(t,\bx)$ such that $\rho(0,\cdot)=\gamma(\cdot)$ and $\lim_{t \to -\infty} \rho(t, \bx) = {\bar \rho} (\bx)$, with ${\bar \rho}$ the stationary profile of  (\ref{eq:pde1}) (resp. (\ref{eq:scl0})). The quasi-potential is the LDF of the empirical density $\rho^N$ in the stationary state $\mu_{\rm{ss}}^N$ of the interacting particle system whose macroscopic behavior is described by (\ref{eq:pde1}) (resp. (\ref{eq:scl0})):
\begin{equation*}
\begin{split}
\mu_{\rm{ss}}^N \Big( \rho^N (\bx) \approx \gamma (\bx) \Big) \approx \exp( -N\nu V^{\,\nu} (\gamma) ), \\
\quad \, \Big({\rm resp.} \quad \mu_{\rm{ss}}^N \Big( \rho^N (\bx) \approx \gamma (\bx) \Big) \approx \exp( -N {\bb V}^{\,\infty} (\gamma) ) \Big).
\end{split}
\end{equation*}
The quasi-potential is usually called entropy or free energy in the physics literature. 
Since $\nu H^\nu$ converges to ${\bb H}^{\infty}$ we have that 
\begin{equation}
\label{eq:qp2}
{\bb V}^{\infty} = \lim_{\nu \to \infty} \nu V^\nu.
\end{equation}
For $\nu>0$, the quasi-potential $V^\nu$ is solution to the Hamilton-Jacobi equation \cite{Gr,FW,MFT}
\begin{equation}
\label{eq:HJ0}
\int_{\RR^2} \left\langle \nabla \cfrac{\delta V^\nu }{\delta \rho} \; , \;  \cfrac{\sigma(\rho)}{2} \, \nabla \cfrac{\delta V^\nu }{\delta \rho} + j_\nu (\rho) \right\rangle \, d\bx =0.
\end{equation}
It follows that $V^\nu$ is a Lyapounov function for the parabolic equation (\ref{eq:pde1}). Indeed we have that if $\rho^\nu$ is solution of (\ref{eq:pde1}), then
\begin{equation*}
\begin{split}
&\frac{d}{dt} \; V^{\nu} (\rho^\nu (t))\; = \int_{\RR^2} \left\langle \cfrac{\delta V^\nu }{\delta \rho} \; , \;  \partial_t  \rho^\nu  \right\rangle \, d\bx =- \int_{\RR^2} \left\langle \cfrac{\delta V^\nu }{\delta \rho} \; , \;  \div \, j_\nu ( \rho^\nu)  \right\rangle \, d\bx\\
& = \int_{\RR^2} \left\langle \nabla \cfrac{\delta V^\nu }{\delta \rho} \; , \;  j_\nu ( \rho^\nu)  \right\rangle \, d\bx = -\cfrac{1}{2}\, \int_{\RR^2} \left\langle \nabla \cfrac{\delta V^\nu }{\delta \rho} \; , \;  \sigma(\rho) \, \nabla \cfrac{\delta V^\nu }{\delta \rho}\right\rangle\, d\bx  \le 0.
\end{split}
\end{equation*}
We recall that we denote by $\rho^{\infty}$ the entropic solution of the conservation law (\ref{eq:scl0}) (or equivalently of (\ref{eq:pde1}) with $\nu=\infty$). It follows from (\ref{eq:qp2}) and from the fact that $\lim_{\nu \to \infty} \rho^\nu=\rho^{\infty}$ that ${\bb V}^{\infty}$ is a Lyapounov function for the entropic solution $\rho^\infty$:
\begin{equation*}
\begin{split}
\frac{d}{dt} \; {\bb V}^{\infty} (\rho^\infty (t)) \le 0.
\end{split}
\end{equation*}

The latter fact can be seen as a form of the second principle. For real physical systems like a gas, described by the laws of the classical mechanics, the scalar conservation law would be replaced by a system of conservation laws (e.g. Euler equations) and the quasipotential, i.e. the entropy, would furnish a non-trivial Lyapounov function of the system. 

The computation of the quasipotential ($V^\nu$ or ${\bb V}^\infty$) is of high interest but usually very difficult to perform. A particular case where an explicit formula is available is when the {\textit{Einstein relation}} holds (see \cite{BFG}):
\begin{equation}
\tag{ER}
D \text{ \; is proportional to \;} \sigma.
\end{equation}
It turns out that in this case the quasipotential is local and take a simple form (see below). If (ER) does not hold, the functional $V^{\nu}$ is non local, i.e. it cannot be written in the form 
$$V^\nu (\gamma) = \int_{\RR^2} v^\nu (\gamma(\bx)) \, d\bx $$
for a suitable function $v^\nu: \RR \to \RR$. Indeed, assume that $V^\nu$ takes this form then by (\ref{eq:HJ0}) we have 
\begin{equation*}
\begin{split}
& \int_{\RR^2} \big[\tfrac{d^2 v^\nu}{d\gamma^2}  (\gamma)  \big]^2 \left \langle \nabla \gamma\, ,\ \Big( \cfrac{\sigma (\gamma)}{2} -\tfrac{1}{\nu \tfrac{d^2 v^\nu}{d\gamma^2}  (\gamma) } D(\gamma) \Big) \nabla \gamma \right\rangle \, d\bx \\
& =\; -\int_{\RR^2} \tfrac{d^2 v^\nu}{d\gamma^2}  (\gamma) \; \langle \nabla \gamma, f (\gamma) \rangle \, d\bx.
\end{split}
\end{equation*}
Since there exists a vector valued function $g: \RR \to \RR^2$ such that $g' = \tfrac{d^2 v^\nu}{d\gamma^2}  \, f$ the RHS of the previous expression is equal to $0$. Since the equality is valid for any profile $\gamma$ we have that $ \nu \tfrac{d^2 v^\nu}{d\gamma^2}(\gamma)   \, \sigma  (\gamma) =2D(\gamma) $. Therefore a necessary and sufficient condition to have a local quasi-potential is that (ER) holds. For Kawasaki dynamics (ER) is satisfied if the dynamics is isotropic but usually not for anisotropic ones. In the context of active particles, (ER) is also rarely satisfied.

It is important to notice that the scalar conservation law (\ref{eq:scl0}) describes only the typical behavior of the microscopic system by forgetting many details of the dynamics. Therefore it describes a priori many different microscopic systems. The quasipotential retains more information of the underlying microscopic dynamics and may be different for microscopic systems whose typical behavior is described by the entropic solution of the same scalar conservation law. 

Let us also notice that in one dimension, the quasipotential associated to a hyperbolic conservation law with Dirichlet boundary conditions has been investigated in \cite{Ba}.

\subsection{Kinetic formulation and associated Entropy production}

A kinetic formulation of the PDE theory of hyperbolic conservation laws has been proposed and developed since \cite{Br, PT, LPT, Pe}. This interpretation will be useful in the sequel. We introduce an auxiliary variable $p\in \RR$ and we define $\chi(p,\rho)={\bf 1}_{0<p \le \rho} -{\bf 1}_{\rho \le p<0}$. Then $\rho$ is a weak solution if, in the sense of distributions, $h(t, \bx,p)=\chi(p, \rho(t, \bx))$ is solution of
\begin{equation}
\label{eq:cin}
\partial_t h + \langle f' (p),  \nabla_{\bx} h \rangle = -\partial_p {\mu_\rho}
\end{equation} 
for some locally finite measure $\mu_{\rho}$ in the form $\mu_\rho (p, dt, d\bx) dp$. To see this, just integrate (\ref{eq:cin}) with respect to $p$ (the RHS becomes $0$) and observe that $\int h(t, \bx, p) dp= \rho(t,\bx)$. In this picture we can imagine that the variable $p$ plays the role of an (artificial) velocity and the measure $\mu_\rho (p, dt, d\bx) dp$ plays the role of the collision term in Boltzmann equation.

Entropy production has a nice form within the kinetic formulation. Assume first that ${\mc V} (p, t,\bx)= \eta (p) \varphi (t,\bx)$ so that  ${\mc Q} (p, t,\bx)=   q(p) \varphi (t,\bx)$ with $q'= f' \eta'$ be the entropy flux associated to $\eta$ and $\varphi$ a smooth test function. Then, by multiplying (\ref{eq:cin}) by $\varphi (t, \bx) \eta' (p)$, integrating in $p$ and using that $h(t, \bx,p)=\chi(p, \rho(t, \bx))$, we get
\begin{equation}
\label{eq:thierry}
\begin{split}
{\mc P}_{\mc V} (\rho)&
= \int \int_\Omega \, \varphi (t,\bx) \eta^{\prime\prime} (p) \mu_\rho (p, dt, d\bx ) \, dp =\int \int_{\Omega} {\mc V} '' (p, t, \bx) \, \mu_{\rho} (p, dt, d\bx ) \, dp.
\end{split}
\end{equation}  
In particular (\ref{eq:thierry}) shows that an entropic solution $\rho$ is such that $\mu_{\rho} (p, \cdot)$ is a negative measure for any $p$. It also shows that
\begin{equation}
\label{eq:entprod}
{\mc P}_{\mc V} (\rho) = \int \, \int_{\Omega} {\mc V} '' (p, t, \bx) \, \mu_{\rho} (p, dt, d\bx ) \, dp
\end{equation}
for an entropy sampler in the form ${\mc V} (p, t,\bx)= \eta (p) \varphi (t,\bx)$. Since a generic entropy sampler ${\mc V}$ can be approximated by a sequence of linear combinations of entropy samplers in the previous form, (\ref{eq:entprod}) is valid for any entropy sampler ${\mc V}$. \\

\begin{figure}
\begin{center}
\begin{tikzpicture}[scale=0.3]
\draw[->,>=latex, Gray] (0,0) -- (15,0);
\draw[->,>=latex, Gray] (0,0) -- (0,15);
\draw[->,>=latex, Gray] (0,0) -- (-13,-10);
\node[below, black] at (15,-9) {\tiny Space};
\node[above, black] at (0,15) {\tiny Time};


\draw [-, line width=1.5 pt, fill, opacity=0.4, shade, bottom color=Gray!80, top color=Gray!10] plot coordinates {(-10,-5) (5,-5) (16,5) (5,5)};
\draw[<-,>=latex, black, line width=0.5 pt] (14,3) -- (17,0) node[right]{\tiny{Plane of constant time $t$}};

\draw [fill, top color=blue!50, bottom color=blue!20, opacity=0.6, shade] plot [smooth cycle] coordinates {(-10,-9) (5,-8) (10,-9) (12,-8) (9,0) (14,11) (4,13) (0,10) (-1,3)};
\draw[<-,>=latex, black, line width=0.5 pt] (13,10) -- (17,13) node[right]{$J_\rho$};

\draw [-, black] plot [smooth] coordinates {(-4,-1) (-2,-1.5) (3,-0.5) (9,-1)};
\draw[black] (3,-0.5) circle (0.7cm);
\node[below, black] at (3,-0.7) {$(t,\bx)$};
\draw[->,>=latex, black, line width=1.5 pt] (3,-0.5) -- (2,0.5);
\node[above, black] at (2,0.5) {$\bn$};
\draw[<-,>=latex, black, line width=0.5 pt] (9.5,-1) -- (17,-5) node[right]{{\tiny{curve}} $s_t$};

\draw[<-,>=latex, black, line width=0.5 pt] (-4,4) -- (1.7,0);
\draw [-, line width=1 pt] plot coordinates {(-5,5) (-5,15) (-15,15) (-15,5) (-5,5)};
\draw [-, line width=1 pt] (-15,9) -- (-5,10);
\draw [->, line width=1 pt] (-10,9.5) -- (-10.2,11);
\node[above,right, black] at (-10.2,11) {$\bn^\bx$};
\node[above,  black] at (-12,12) {$\rho^+$};
\node[above, black] at (-8,6) {$\rho^-$};

\end{tikzpicture}
\end{center}
\caption{Local structure of a weak solution.}
\label{fig:2} 
\end{figure}
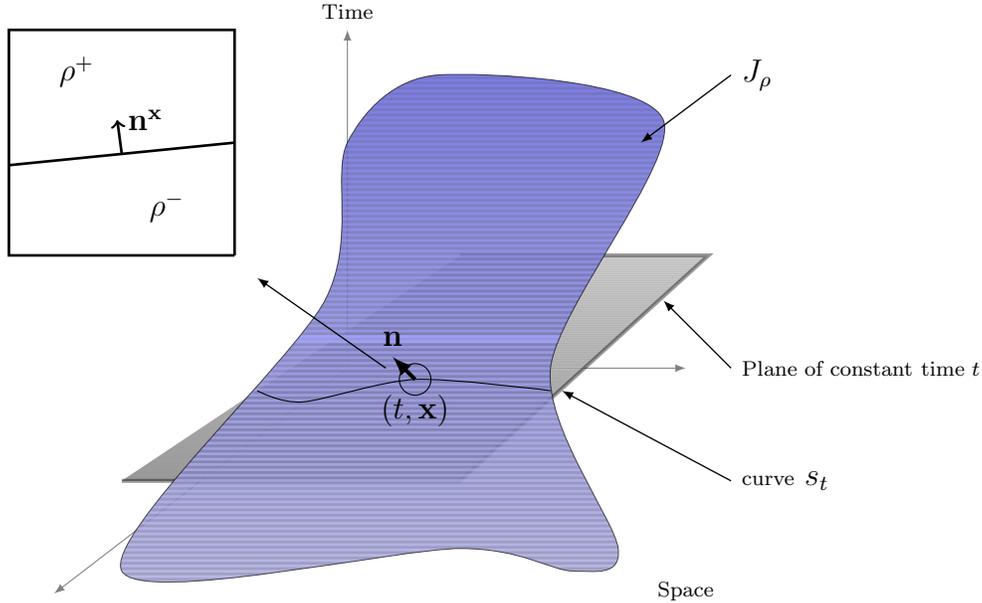

\subsection{Explicit formula }
\label{sec:ep0}

Our aim is now to give a more explicit formula for $\mu_\rho$ and thus for ${\mc P}_{\mc V} (\rho)$. Let $\rho(t,\bx)$ be a weak solution and denote by $J:=J_\rho$ its jumps set that we assume to be such that for any $t \in [0,T]$, $s_t$ is a smooth curve {\footnote{The regularity of the jump sets of weak solutions is studied in \cite{LOW}. It is not smooth in general but is sufficiently regular to define almost everywhere a unit normal to the jumps set.}} $\alpha\in[0,1] \to s_t (\alpha) \in \RR^2$. Then we have that $J= \cup_{t \in [0,T]} \, s_t =\{ (t, s_t (\alpha)) \} \subset \Omega$ is a $2$-dimensional manifold parameterized by $(t,\alpha)$. Let $\Omega^{\pm}$ be the two connected components of $\Omega \backslash J$.  A unit normal vector to $J$ at $(t,\bx)=(t, s_t (\alpha)) \in J$ is 
\[
\bn:=(\bn^t, \bn^\bx)= \tfrac{1}{\mathcal{N}}\; \left(- \left\langle \tfrac{ds_t}{dt}, \big[\tfrac{d s_t}{d\alpha}\big]^\perp \right\rangle, \big[\tfrac{d s_t}{d\alpha}\big]^\perp\right)\in \RR^3,
\]
with $\mathcal{N}:={\mc N}_{(t,\bx)}$ a normalization factor.
For any $(t,\bx)\in J$ let $\rho^\pm :=\rho^\pm (t, \bx)$ be the limit of $\rho(s, \by)$ as $(s, \by) \in \Omega^{\pm} \to (t,\bx)$. Observe that $\rho$ is regular in $\Omega^{+} \cup \Omega^{-}$ and discontinuous on $J$. See Figure \ref{fig:2}. An integration by parts and Green's theorem shows that if $\varphi$ is a smooth function vanishing at time $0$ and time $T$,
\begin{equation}
\label{eq:lkjh}
\begin{split}
&- \int_{\Omega} \eta (\rho) \partial_t \varphi \; dt d\bx - \int_\Omega \langle q(\rho), \nabla \varphi \rangle \; dt d\bx \\
&= \int_{\Omega^-}  \left[ \partial_t \eta (\rho) + \div q(\rho) \right] \varphi \;dt d\bx +\int_{\Omega^+}  \left[ \partial_t \eta (\rho) + \div q(\rho) \right] \varphi \; dt d\bx \\
&+ \int_J \left\{ (\eta^+ -\eta^-) \bn^t + \langle (q^+ -q^-),  \bn^{\bx} \rangle \right\} \varphi \; d\gamma_J
\end{split}
\end{equation}
where $d\gamma_J={\mc N}_{(t,\bx)} dt d\alpha$ is the Lebesgue measure on $J$ and $\eta^\pm=\eta(\rho^\pm)$, $q^{\pm} =q(\rho^\pm)$. Since $\rho$ is regular in $\Omega^{\pm}$ the two first integrals on the RHS of the last equality of (\ref{eq:lkjh}) are zero and we conclude by (\ref{eq:thierry}) that, as space-time measures, 
\begin{equation}
\label{eq:tre}
\int dp \, \eta'' (p) \mu_\rho (p, dt,d\bx) = {\bf 1}_{(t, \bx) \in J}\;  \left\{ (\eta^+ -\eta^-) \bn^t + \langle (q^+ -q^-),  \bn^{\bx} \rangle \right\}\; d\gamma_J.
\end{equation}
Remark that up to now we did not use any convexity property of $\eta$ and that the choice $\eta (v) =v, q(v) =f(v)$ is valid. Since $\rho$ is a weak solution, this choice implies that the LHS of (\ref{eq:lkjh}) is then $0$ and this gives the {\textit{Rankine-Hugoniot condition}}:
\begin{equation}
\label{eq:RH}
 (\rho^+ -\rho^-) \bn^t + \langle (f (\rho^+)  - f(\rho^-) ),  \bn^{\bx} \rangle =0
\end{equation}
on $J$. Now we write
$$\eta^+ -\eta^- = \int_{\rho^-}^{\rho^+} \eta{'} (z) dz = - \int_{\rho^-}^{\rho^+} (z- \rho^-) \eta{''} (z) dz + (\rho^+ -\rho^-) \eta{'} (\rho^+)$$
and
\begin{equation*}
\begin{split}
q^+ -q^- &= \int_{\rho^-}^{\rho^+} q{'} (z) dz= \int_{\rho^-}^{\rho^+} \eta' (z) f' (z) dz\\
&= - \int_{\rho^-}^{\rho^+} (f(z)- f(\rho^-)) \eta{''} (z) dz + (f(\rho^+) -f(\rho^-)) \eta{'} (\rho^+).
\end{split}
\end{equation*}
We plug this in (\ref{eq:tre}) and use the Rankine-Hugoniot condition (\ref{eq:RH}) to get 
\begin{equation}
\begin{split}
\int dp \, \eta'' (p) \mu_\rho (p, dt,d\bx) &= {\bf 1}_{(t, \bx) \in J}\; \left\{ - \int_{\rho^-}^{\rho^+} \eta'' (p) (p-\rho^-) \bn^t dp \right.\\
&\left. - \int_{\rho^-}^{\rho^+} \eta'' (p) \langle (f(p) -f(\rho^-)), \bn^\bx \rangle dp  \right\} \; d\gamma_J.
\end{split}
\label{eq:muvu2}
\end{equation} 
Since this is true for any sufficiently regular $\eta$ we get (see Appendix \ref{sec:appendixA})
\begin{equation}
\label{eq:muvu}
\mu_\rho (p, dt, d\bx) =  {\bf 1}_{(t, \bx) \in J}\;  {\theta} (p,\rho^-, \rho^+) \; d\gamma_J
\end{equation}
with  
\begin{equation}
\label{eq:thetachaise}
\theta (p,\rho^-,\rho^+) = {\bf 1}_{ \inf(\rho^-,\rho^+) \le p \le \sup(\rho^-, \rho^+)}  \; \Gamma_{\tiny{\tfrac{{\bf n}^{\bx}}{\| {\bf n}^\bx \|}}} (p, \rho^-,\rho^+) \; \| {\bf n}^\bx\|
\end{equation}
where $\Gamma_{\bk} (p, \rho^-,\rho^+)$, $\|\bk \|=1$ is defined by
\begin{equation}
\label{eq:thetachaisegamma}
\Gamma_{\bk} (p,\rho^-,\rho^+)= \cfrac{ \; \left\langle f(\rho^-) (\rho^+ -p) + f(\rho^+) (p-\rho^-) -f(p) (\rho^+ -\rho^-) \, , \, \bk \right\rangle}{|\rho^+ -\rho^-|}.
\end{equation}
Observe that $\| {\bf n}^\bx \| d\gamma_J=\| \big[\tfrac{d s_t}{d\alpha}\big]^\perp \| d\alpha\, dt=\| \tfrac{d s_t}{d\alpha} \| d\alpha\, dt= dt \, ds_t $ where $ds_t$ is the Lebesgue measure on the curve $s_t$.

\subsection{Entropy splittable solution, shocks and anti shocks}

From \eqref{eq:muvu}, we see that the measure $\mu_\rho$ is concentrated on the jump set $J_\rho$ of the weak solution $\rho$.
For a general weak solution we denote by 
\begin{equation}
\label{eq:mu_rho_pm}
\begin{split}
&\mu_\rho^\pm (p, dt, d\bx)={\bf 1}_{(t, \bx) \in J}\;  {\theta}^{\pm} (p,\rho^-, \rho^+) \; d\gamma_J\\
&={\bf 1}_{(t, \bx) \in J} {\bf 1}_{ \inf(\rho^-,\rho^+) \le p \le \sup(\rho^-, \rho^+)}  \; \Gamma^{\pm}_{\tiny{\tfrac{{\bf n}^{\bx}}{\| {\bf n}^\bx \|}}} (p, \rho^-,\rho^+) \; dt \, d s_t
\end{split}
\end{equation}
the (time-space) positive and negative parts of the measure $\mu_\rho(p, \cdot)$ and by $E^{\pm}_\rho (p) \subset J_\rho$ their support. Observe that roughly $E^{\pm}_\rho (p)$ is the set of $(t,\bx)\in J_\rho$ for which entropy is produced (resp. dissipated) by the weak solution $\rho$ when $\rho(t,\bx)=p$.\\

Let us first state precise definitions of \emph{shocks} and \emph{antishocks}:
\begin{definition}
 Let $\rho(t,\bx)$ be a weak solution.  A point $(t,\bx)\in \Omega$ is said to be a \emph{shock} if it always dissipate entropy:
\[
(t,\bx) \in \cap_p E_\rho^- (p) ;
\] 
it is said to be an \emph{antishock} if it may produce entropy:
\[
(t,\bx) \in \cup_p E_\rho^+ (p).
\] 
\end{definition}

Observe that a point $(t, \bx) \in J_\rho$ may belong to $E_\rho^+(p)$ and at same time belong to $E_\rho^- (p^\prime)$ for some $p \ne p^{\prime}$. Following \cite{BBMN}, we now introduce a special class of weak solutions for which this does not happen.

\begin{definition}
A weak solution $\rho(t,\bx)$ is said to be \emph{entropy splittable} if for any discontinuity, characterized by $\rho^-,\rho^+$ and a local unit vector $\bk$, 
the quantity  $\Gamma_{\bk} (p,\rho^-,\rho^+)$ has a constant sign when $p$ varies in $[\rho^-,\rho^+]$. In other words,
\[
\cup_p E_\rho^+ (p) \cap \cup_p E_\rho^- (p) = \varnothing.
\]
\end{definition}

The entropy splittable weak solutions will play a technical role later on. 
Entropy splittable solutions only have shocks and ``perfect antishocks", that always produce entropy (ie points $(t,\bx) \in \cap_p E_\rho^+ (p)$).

\begin{remark}
In dimension $1$, if the flux function $f$ is convex or concave, any weak solution is entropy splittable. 
\end{remark}

\section{Main Results: Density Large Deviation Function}
\label{sec:densityLDF}

We turn now to the main object of our study, the large deviation function for a density profile $\rho$. The density LDF of the considered weakly drifted interacting particle system is given by
\begin{equation*}
H_{[0,T]}^{\nu} (\rho)= \inf_{j} {\mc I}^\nu_{[0,T]} (j, \rho)
\end{equation*}
where the infimum is carried over currents $j$ satisfying the constraint 
$$\partial_t \rho + \div j =0 .$$
We are interested in the behavior of $\nu H^{\nu}_{[0,T]}$ as $\nu \to \infty$. It is conjectured that $\nu {H}^{\nu}$ converges{\footnote{More exactly, $\Gamma$-converges.  }} to a functional ${\bb H}_{[0,T]}^{\infty}$ which is the LDF of the empirical density for the strongly drifted underlying system of interacting particles.\\

In this section we argue that in the $2d$-case, the functional ${\bb H}_{[0,T]}^{\infty}$ is 
\begin{equation}
\label{eq:conj}
{\bb H}^{\infty}_{[0,T]} (\rho) = 2 \int_0^T \, \int_{\bx \in s_t}  \, \left\{ \int_{\rho^- \wedge \rho^+ }^{\rho^- \vee\rho^+} \cfrac{\langle \bn^{\bx}, D(p) \, \bn^{\bx} \rangle}{\langle {\bn^\bx}, \sigma(p) \, {\bn^{\bx}} \rangle } \; \Gamma_{\tfrac{\bn^\bx}{\|\bn^\bx\|}}^+ (p, \rho^-,\rho^+)\, dp\right\} \, dt \, ds_t
\end{equation}
where $\Gamma^+$ is the positive part of the function $\Gamma$ which is defined by (\ref{eq:thetachaisegamma}) and $ds_t$ is the Lebesgue measure on $s_t$. The notations are those of Figure~\ref{fig:2} and Section \ref{sec:ep0}.

\subsection{Large deviation function lower bounds in terms of the entropy production}
\label{sec:lowerbounds}
The aim of this section is to prove the following lower bound for $\bb H_{[0,T]}^\infty$ 
\begin{equation}
\label{eq:conj2}
{\bb H}^{\infty}_{[0,T]} (\rho) \ge 2 \int_0^T \, \int_{\bx \in s_t}  \, \left\{ \int_{\rho^- \wedge \rho^+ }^{\rho^- \vee\rho^+} \cfrac{\langle \bn^{\bx}, D(p) \, \bn^{\bx} \rangle}{\langle {\bn^\bx}, \sigma(p) \, {\bn^{\bx}} \rangle } \; \Gamma_{\tfrac{\bn^\bx}{\|\bn^\bx\|}}^+ (p, \rho^-,\rho^+)\, dp\right\} \, dt \, ds_t.
\end{equation}

In order to prove it we first show a non-optimal lower bound whose derivation will be however useful for our purpose. This non-optimal lower bound is the following:  for ``any" weak solution $\rho(t,{\bx})$ of the scalar conservation (\ref{eq:scl0})
\begin{equation}
\label{eq:liminf0}
\bb H_{[0,T]}^\infty (\rho)  \ge \sup_{{\mc V} \in {\bf {\hat V}}} {\mc P}_{\mc V} (\rho).
\end{equation}  
We recall that the entropy production ${\mc P}_{\mc V} (\rho)$ has been defined in (\ref{eq:entprod}) and we denote by ${\bf {\hat V}}$ the set of convex entropy samplers (${\mc V}'' \ge 0$, where $'$ denotes the derivative with respect to the first argument) satisfying the relaxed Einstein condition
 \begin{equation}
 \label{eq:relEin}
 2 D(p) \ge \sigma(p) {\mc V}^{\prime\prime} (p, t, \bx)
 \end{equation}
for any $(p,t, {\bx}) \in \mathbb{R} \times \Omega$.\\

More precisely, by definition of the $\Gamma$-convergence (see footnote \ref{foot:gamma}), in order to show this lower bound we have to prove that 
\begin{equation}
\label{eq:liminf}
\liminf_{\nu \to \infty}\;  \nu {H}^{\nu} (\tilde \rho^{\nu}) \ge \sup_{{\mc V} \in {\bf {\hat V}}} {\mc P}_{\mc V} (\rho)
\end{equation}  
for any sequence of smooth functions $\tilde \rho^{\nu} (t, \bx)$ converging to the weak solution $\rho (t, {\bx})$ in a suitable topology. This proof is a simple extension \footnote{We thank C. Bahadoran for having brought this to our attention.} of the proofs of Theorem 2.5, (i) of \cite{BBMN} or Theorem 2.1 in \cite{BM}. 
To prove (\ref{eq:liminf}) we will need the following dual variational characterization of $H^\nu$ which is proved in Appendix \ref{sec:app-vf}:
\begin{equation}
\label{eq:varfl}
{\nu}H^\nu (u)= \sup_{\varphi} \left\{ \ell_{\nu,u}(\varphi)-\frac{1}{2\nu} \int  \langle \nabla \varphi,\sigma \nabla \varphi \rangle \, dt d\bx\right\}
\end{equation}
with
\[
\ell_{\nu,u}(\varphi) = \int  \big(\partial_t u + \div j_\nu ( u)\big)\, \varphi  \, dt d\bx  .
\]

Let $({\mc V},{\mc Q})$ be an entropy- entropy flux couple sampler, and let $\varphi^\nu (t,\bx) = {\mc V}'(\tilde \rho^\nu(t,\bx),t,\bx))$ (we recall that $'$ denotes the derivative with respect to the $\rho$ argument and $\nabla_\bx$ the gradient with respect to the space variable).
Then 
\[
\nabla \varphi^\nu = \nabla_\bx {\mc V}' +{\mc V}" \nabla \tilde \rho^\nu.
\]
Thus
\begin{equation}
\label{eq:ll9991}
\begin{split}
&\nu H^{\nu} (\tilde \rho^{\nu}) \ge  \ell_{\nu,\tilde \rho^\nu}(\varphi^\nu)-\frac{1}{2\nu} \int \langle \nabla \varphi^\nu,\sigma \nabla \varphi^\nu \rangle \, dtd\bx \\
&= \int \big[\partial_t \tilde \rho^\nu +  \div  f(\tilde \rho^\nu) \big] {\mc V}' \, dt d\bx  +\frac{1}{\nu} \int \langle D(\tilde \rho^\nu)\nabla \tilde \rho^\nu,{\mc V}" \nabla \tilde \rho^\nu \rangle \, dt d\bx  \\
&  +\frac{1}{\nu} \int  \langle D(\tilde \rho^\nu)\nabla \tilde \rho^\nu,\nabla_\bx {\mc V}'\rangle \, dt d\bx  - \frac{1}{2\nu} \int \langle \nabla_\bx {\mc V}',\sigma (\tilde \rho^\nu)  \nabla_\bx {\mc V}' \rangle\,  dt d\bx  \\
& - \frac{1}{\nu} \int \langle \nabla_\bx {\mc V}',\sigma (\tilde \rho^\nu) {\mc V}" \nabla \tilde \rho^\nu\rangle \, dt d\bx  -\frac{1}{2\nu}  \int \langle {\mc V}" \nabla \tilde \rho^\nu,\sigma (\tilde \rho^\nu)  {\mc V}" \nabla \tilde \rho^\nu\rangle \, dt d\bx  \\
&= -\int \left[(\partial_t {\mc V})(\tilde \rho^\nu(t,\bx),t,\bx)+(\nabla_\bx{\mc Q})(\tilde \rho^\nu(t,\bx),t,\bx)\right] \, dt d\bx  \\
&+\frac{1}{2\nu} \int\langle [2D(\tilde \rho^\nu)-{\mc V}"\sigma(\tilde \rho^\nu)]\nabla \tilde \rho^\nu,{\mc V}" \nabla \tilde \rho^\nu\rangle \, dt d\bx\\
&+\frac{1}{\nu} \int  \langle D(\tilde \rho^\nu)\nabla \tilde \rho^\nu,\nabla_\bx {\mc V}'\rangle \, dt d\bx  \\
& - \frac{1}{\nu} \int \langle \nabla_\bx {\mc V}',\sigma {\mc V}" \nabla \tilde \rho^\nu\rangle \, dt d\bx 
- \frac{1}{2\nu} \int \langle \nabla_\bx {\mc V}',\sigma \nabla_\bx {\mc V}'\rangle \, dt d\bx.
\end{split}
\end{equation}

Apply first this inequality in the particular case $(\mc V (p, t,\bx) , \mc Q(p, t,\bx))=(c p^2,cf(p) )$ and for some constant $c>0$ sufficiently small to ensure that $2D(p) -2c \sigma(p) \ge \kappa >0$.  Then, the last RHS of (\ref{eq:ll9991}) contains only one non zero term and we get  
$$\nu H^{\nu} (\tilde \rho^{\nu}) \ge \cfrac{c \kappa}{\nu} \int\langle \nabla \tilde \rho^\nu, \nabla \tilde \rho^\nu\rangle \, dt d\bx.$$
Therefore we can always assume (otherwise extract a subsequence) that the approximation $\tilde \rho^\nu$ satisfies
$$ \sup_{\nu >0} \cfrac{1}{\nu} \int\langle \nabla \tilde \rho^\nu, \nabla \tilde \rho^\nu\rangle \, dt d\bx < \infty.$$

Now, the last three terms in (\ref{eq:ll9991}) vanish in the $\nu \to \infty$ limit, because they contain strictly less than two $\nabla \tilde \rho^\nu$ (apply Cauchy-Schwarz inequality and use the previous bound). The first term in the RHS of the last equality of (\ref{eq:ll9991}) is exactly
${\mc P}_{\mc V} (\tilde \rho^\nu)$, which tends to ${\mc P}_{\mc V} (\rho)$. Furthermore, if ${\mc V}$ is chosen such that
\begin{equation}
\label{ineq}
\langle [2D(\tilde \rho^\nu)-{\mc V}"\sigma(\tilde \rho^\nu)]\nabla \tilde \rho^\nu,{\mc V}" \nabla \tilde \rho^\nu\rangle  \geq 0
\end{equation}
then the second term is positive. One possibility is to choose ${\mc V}$ convex (w.r.t. $\rho$), and satisfying (\ref{eq:relEin}); this leads to the bound \eqref{eq:liminf}. \\

Let us now explain why this lower bound is not optimal and how it can be improved. Let us recall that around any $(t,\bx)\in J_\rho$, the weak solution looks like a step function propagating in the direction of the unit vector $\bn(t,\bx)$ with a velocity prescribed by the Rankine-Hugoniot condition (see Figure \ref{fig:2}). Moreover $\nabla \tilde \rho^\nu$ is actually parallel to $\bn (t, \bx)$, up to subdominant terms. Thus, if we repeat the previous computations but with a convex entropy samplers ${\mc V}$ such that
\begin{equation}
{\mc V}"(p,t,\bx) \big\langle \sigma(p) \bn(t,\bx)\, , \, \bn(t,\bx) \big\rangle \; \leq 2 \big\langle D(p)\bn(t,\bx)\, , \, \bn(t,\bx) \big\rangle,
\label{ineq:Vrho}
\end{equation}
we will still have (because $\nabla \tilde \rho^\nu$ is almost parallel to $\bn (t, \bx)$) 
\begin{equation}
\label{eq:liminf_gen}
\liminf_{\nu \to \infty} \; \nu {H}^{\nu} (\rho_{\nu}) \ge  {\mc P}_{\mc V} (\rho).
\end{equation}
Let $\bf{\hat V}_{\rho}$ be the set of convex entropy samplers ${\mc V}$ such that for any $(t,\bx) \in J_\rho$ the inequality (\ref{ineq:Vrho}) is satisfied. We deduce that
\begin{equation*}
\bb H_{[0,T]}^\infty (\rho)  \ge \sup_{{\mc V} \in {\bf {\hat V}}_\rho} {\mc P}_{\mc V} (\rho).
\end{equation*}  
We now choose ${\mc V}(p,t,\bx)=0$ if $(t,\bx)\in E_\rho^-(p)$ (an "entropic point" does not cost anything), 
and
\[
{\mc V}''(p,t,\bx)= \frac{2\langle D(p)\bn(t,\bx), \bn(t,\bx) \rangle}{\langle \sigma(p) \bn(t,\bx),\bn(t,\bx)\rangle}~\mbox{\; if \; }~(t,\bx)\in E_\rho^+(p).
\]
Such a choice may violate the regularity requirements for an entropy sampler; in this case, it is necessary to consider a regularization of the above choice, introducing serious mathematical complications that we disregard. Using \eqref{eq:entprod} and \eqref{eq:mu_rho_pm}, we see that the RHS of \eqref{eq:liminf_gen} coincides with the RHS of \eqref{eq:conj2}, which is then proved. Note that since $\bf{\hat V}_{\rho}$ is in general a larger set than $\bf{\hat V}$, the lower bound \eqref{eq:liminf0} is in general not optimal.


\subsection{The one-dimensional generalized Jensen-Varadhan functional and the additive principle}

We have proved in the previous section the lower bound part of the conjecture (\ref{eq:conj}). We now review the results obtained in the one-dimensional case in \cite{J,V} (for a particular flux) and in \cite{BBMN} (for a general non convex hyperbolic flux). This will further substantiate conjecture (\ref{eq:conj}), and emphasize what is missing to prove it.





\subsubsection{The one-dimensional generalized Jensen-Varadhan functional}
In \cite{BBMN} it is proved rigorously\footnote{In fact, in \cite {BBMN}, the $\Gamma$-convergence of $\nu H^\nu (\rho)$ to ${\bb H}^\infty _{[0,T]} (\rho)$ is only proved for ``entropy splittable" weak solutions and the extension to generic weak solution would require a density argument which appears as a very difficult problem (see the comments after Theorem 2.5 there).} 
that conjecture (\ref{eq:conj}), simplified according to the fact that $D$ and $\sigma$ are scalars, is correct; hence, with $\Omega=[0,T] \times \RR$ 
\begin{equation}
\label{eq:hinfnurho}
{\bb H}^{\infty}_{[0,T]} (\rho) = 2 \int \cfrac{D(p)}{\sigma (p)}  \; {\mu}_{\rho}^+ (p \, , \, E_\rho^+ (p) ) \, dp = 2 \int \cfrac{D(p)}{\sigma (p)}  \; {\mu}_{\rho}^+ (p \, , \, \Omega) \, dp.
\end{equation}
Notice that \eqref{eq:hinfnurho} is precisely the one dimensional equivalent of \eqref{eq:conj} when $D$ and $\sigma$ are scalars. This can be seen by replacing ${\mu}_{\rho}^+ (p \, , \, \Omega)$ by its explicit expression given by the one dimensional versions of \eqref{eq:muvu},\eqref{eq:thetachaise},\eqref{eq:thetachaisegamma}. In particular, ${\mu}_{\rho}^+ (p \, , \, \Omega)$ implicitly contains an integral over the jump set of the weak solution $\rho$.

We call this functional the {\textit{generalized Jensen-Varadhan}} functional. It is important to notice that the only points $(t,x) \in \Omega$ which contribute in ${\bb H}_{[0,T]}^{\infty}(\rho)$ are points in $J_\rho$ which produce entropy, i.e. the anti-shocks. 
The integral formula (\ref{eq:hinfnurho}) shows that all the contributions of anti-shocks simply add up: we now make this remark more precise.

\subsubsection{Case of a step function in one dimension}

First, consider the particular case where $\rho$ is a step function separated by a discontinuity moving at velocity $v$ and taking the values $\rho(t,x)=\rho^-$ for $x -tv \le x_*$, $\rho(t,x) =\rho^+$ for $x-tv>x_*$. We assume $\rho^- <\rho^+$ and let $f_\pm=f(\rho^\pm)$. Observe that the velocity $v$ has to be equal to $(f(\rho^+) -f(\rho^-))/(\rho^+  -\rho^-)$ to ensure that $\rho$ is a weak solution (Rankine-Hugoniot condition, see (\ref{eq:RH}) ). We have $J_\rho=\{ (t, x_*+tv)\, ;\, t\in [0,T] \}$ where $x_* +tv$ is the position of the discontinuity at time $t$. Let $\Gamma$ be the one-dimensional version of the function $\Gamma$ introduced in (\ref{eq:thetachaisegamma}). If $f$ is convex or concave on $[\rho^-,\rho^+]$ then $\Gamma(\cdot, \rho^-,\rho^+)$ has a constant sign, i.e. $\rho$ is entropy splittable, so that the discontinuity corresponds either to a shock or to a ``perfect" antishock in the sense that it belongs to $\cap_{p} E_\rho^+ (p)$. Then by (\ref{eq:muvu}), (\ref{eq:hinfnurho}) becomes (see also Remark 2.7 in \cite{BBMN})
\begin{equation}
\label{eq:hinfnurho2}
{\bb H}_{[0,T]}^{\infty} (\rho)=2 {\frac{|J_\rho|}{\sqrt{1+v^2}}} \int_{\rho^-}^{\rho^+} \cfrac{D(p)}{\sigma (p)}\;  | \Gamma (p, \rho_-,\rho_+)|\, dp
\end{equation} 
if $\rho$ is not an entropic solution and $0$ if it is. Here $|J_\rho|=T\sqrt{1+v^2}$ is the length of the discontinuity set on $[0,T]$. As noticed in \cite{BD}, the term 
$\tfrac{H_{[0,T]}^{\infty} (\rho)}{T}$
coincides with the cost to produce a time averaged current equal to $i_0= x_* f_- +(1-x_*) f_+$ in the large $\nu$ limit. In the case where the flux is neither convex, nor concave, the function $\Gamma(\cdot,\rho^-,\rho^+)$ may change sign on the interval $[\rho^-,\rho^+]$; in this case, the discontinuity may be an anti-shock but it is not a ``perfect antishock": the solution is not ``entropy splittable". Then, the term $|\Gamma|$ in (\ref {eq:hinfnurho2}) has to be replaced by $\Gamma^+$ \cite{M}.
 From a mathematical point of view, all this has been rigorously derived by a LDP (\cite{V}) for the $1d$ asymmetric simple exclusion process, which corresponds to the case $f(u)=\sigma (u) =u(1-u)$ and $D(u)=1$.
 
 \subsubsection{Additive principle} 
 \label{sec:additive_principle}

It is interesting to notice that the formula (\ref{eq:hinfnurho2}) is in fact sufficient to recover (\ref{eq:hinfnurho}) by assuming a (space-time)  additive principle \cite{BD2}: \\

{\textit{{\underline{Additive principle:}} For a generic weak solution only antishocks contribute to ${\bb H}_{[0,T]}^{\infty}$ and simply add up. Moreover, the contribution of each antishock can be evaluated by approximating locally the weak solution by a moving step function.}}     \\


In the $2d$-case, it is tempting to follow this route: solve the problem for a simple step profile; then use the additive principle to obtain an
expression for ${\bb H}^\infty$ for a general profile. This is our aim in the following section.

\subsection{Case of a moving step function}
\label{sec:conjecture_checking_step}

We consider a moving step profile $\rho$  between $\rho^-$ and $\rho^+$ moving in some arbitrary direction ${\bk} \in \RR^2$, $\| \bk \|=1$, with some velocity $v \in \RR^2$. The RHS  of (\ref{eq:conj}) is then infinite since the jumps set of $\rho$ is a strip of $\RR^2$ and that the RHS is roughly extensive in the area of the jump set. Therefore our aim will consist to evaluate 
\begin{equation*}
\lim_{L\to\infty} \tfrac{1}{2L} \; \lim_{\nu \to \infty} \nu H_{[0,T]}^{\nu, L} (\rho)
\end{equation*}
where to define $H_{[0,T]}^{\nu, L} (\rho)$ we replace in (\ref{eq:Hnunu}) the set $\Omega=[0,T]\times \RR^2$ by the set $\Omega_L= [0,T]\times [-L,L]^2$. We then show that the previous limit is equal to
\begin{equation}
\label{eq:JVgen_movingstep}
{\bar H} (\rho) = 2T \;\int_{\rho^- \wedge \rho^+ }^{\rho^- \vee\rho^+} \,  \cfrac{\langle {\bk} , D(p) \, \bk \rangle}{\langle {\bk}, \sigma(p) \, {\bk} \rangle } \; \Gamma_{\bk}^+ (p,\rho^-,\rho^+) \, dp. 
\end{equation}
To simplify the notation we omit the index $[0,T]$.

The lower bound does not follow directly from (\ref{sec:lowerbounds}), since the RHS of \eqref{eq:conj} is infinite for a step function; we then actually have to repeat the arguments of Section \ref{sec:lowerbounds}, using samplers supported in $\Omega_L$, see Section \ref{sec:lbmsf}. Also, serious mathematical difficulties have been disregarded in Section \ref{sec:lowerbounds}; by contrast, the arguments presented below for a step function in Section \ref{sec:lbmsf} are essentially rigorous.
We first derive the upper bound{\footnote{In the sense of the $\Gamma$-convergence, see footnote \ref{foot:gamma}.}} by obtaining a smooth approximation $\tilde \rho^\nu$ of $\rho$ such that
$$\lim_{L\to\infty} \tfrac{1}{L} \;  \lim_{\nu \to \infty} {\nu} {H}^{\nu, L} (\tilde \rho^\nu) \; {=}\;  {\bar H} (\rho)$$
To be more precise, following the strategy of \cite{BBMN}, we first prove the upper bound of \eqref{eq:JVgen_movingstep} for an ``entropy splittable" moving step. We then argue (in a not fully rigorous manner) that this upper bound holds for general moving steps.

For an entropic moving step, \eqref{eq:JVgen_movingstep} is obviously valid, since both LHS and RHS vanish. For a non entropic entropy splittable moving step, the RHS of \eqref{eq:JVgen_movingstep} can be rewritten
\begin{equation}
\begin{split}
2 T \, \int_{\rho^- \wedge \rho^+ }^{\rho^- \vee\rho^+}\,  \cfrac{\langle D(p) \bk, \bk\rangle}{\langle \sigma (p) \bk, \bk \rangle} \,  |\Gamma_{\bk} (p, \rho^-, \rho^+)| \, dp.
\end{split}
\label{eq:JVgen_es}
\end{equation}
This is the upper bound we shall show in the next section.

\subsubsection{Upper bound for an entropy splittable step}
\label{UB_entrop_split}

A weak solution of (\ref{eq:scl0}) in the form of a moving step profile $\rho$ between $\rho^-$ and $\rho^+$ moving in some arbitrary direction ${\bk} \in \RR^2$, $\| \bk \|=1$, with velocity $v$, is given, and we want to approximate it in an optimal way. Let us look for a traveling wave propagating with velocity {$v$} in the direction $\bk$ given by
$$\rho^{\nu} (t, \bx) =g^{\nu} (\langle \bk, \bx \rangle -{v} t)$$
solution of
\begin{equation*}
\partial_t \rho^\nu + \div f(\rho^\nu) = {\frac{1}{\nu}} \, \div ( {D}(\rho^\nu) \nabla \rho^\nu)
\end{equation*}
on $\RR^2$ with boundary conditions 
\begin{equation}
\label{eq:bc567}
\lim_{z \to \pm \infty} g^\nu (z)= \rho^\pm.
\end{equation}
A simple computation shows that $g^\nu$ shall satisfy for any $z\in \RR$ that
\begin{equation}
\label{eq:5hmat}
\langle \bk, {D}(g^\nu (z))  \bk \rangle\,  \cfrac{d g^{\nu}}{dz} (z)  = \nu \,\big[  \langle \bk, f(g^\nu(z)) \rangle -v g^\nu (z) +C \big]
\end{equation}
for some constant $C:=C(\bk,\nu)$. Since this equation is invariant by $z \to z+c$ for any constant $c$, in order to fix uniquely $g^\nu$, we impose that $g^\nu (0) =(\rho^- + \rho^+)/2$. We then consider the restriction of $\rho^\nu$ on $[-L,L]^2$, and introduce $\tilde \rho^{\nu}$ as the time reversal of $\rho^{\nu}$: 
\begin{equation}
\tilde \rho^{\nu} (t, \bx) = \rho^{\nu} (-t, -\bx).
\end{equation}
We have that ${g}^{\nu} (t, \bx)$ converges to an entropic solution $g^{\infty} ( \langle \bk, \bx\rangle - vt)$ with $g^{\infty}$ a step function with a shock located at $z_s=0 \in \RR^2$:
\begin{equation*}
g^{\infty} (z) = \rho^- {\bf 1}_{z<z_s} +\rho^{+} {\bf 1}_{z \ge z_s}.
\end{equation*}
Then $\tilde \rho^\nu(t, \bx)$ converges as $\nu \to \infty$ to a non-entropic moving step profile $\tilde \rho^{\infty} (t,\bx)=g^{\infty} ( -\langle \bk, \bx\rangle +vt)$. Observe that the anti-shock of $\tilde \rho^\nu$ is present in $\bx \in [-L,L]^2$ at time $t \in [0,T]$ when 
\begin{equation}
z_s = -\langle \bk, \bx \rangle +vt.
\end{equation} 
Otherwise it is not. By (\ref{eq:5hmat}) and (\ref{eq:bc567}), we have that
\begin{equation}
\label{eq:cinfty}
\begin{split}
&C_\infty:=\lim_{\nu \to \infty} C(\bk, \nu)= v \rho^- -\langle f(\rho_-), \bk \rangle = v \rho^+ - \langle f(\rho^+), \bk \rangle,\\
&v=\left\langle \cfrac{f(\rho^+)-f(\rho^-)}{\rho^+ -\rho^-}, \; \bk \right\rangle.
\end{split}
\end{equation}
The second condition is nothing but the Rankine-Hugoniot condition that we could have assumed ab initio. Since  
\begin{equation}
\partial_t \tilde \rho^\nu = - \div {\tilde j}_\nu  (\tilde \rho^\nu), \quad {\tilde j}_\nu (\tilde \rho^\nu) = f(\tilde \rho^\nu) + \frac{1}{\nu}{D} (\tilde \rho^\nu) \nabla \tilde \rho^\nu
\end{equation}
any $j$ such that $\partial_t \tilde \rho^\nu = -\div j$ is in the form 
\begin{equation}
\label{eq:currentj}
j = {\tilde j}_\nu(\tilde \rho^\nu) + \frac{1}{\nu} \nabla^{\perp} \varphi
\end{equation}
for some function $\varphi$. We have then
\begin{equation}
\begin{split}
&\nu {H}^{\nu,L} (\tilde \rho^\nu)\\
&=\cfrac{1}{4 \nu L}  \inf_{\varphi} \left\{  \int_{\Omega_L} \, \left\langle  \left[ \nabla^{\perp} \varphi + {2 D}(\tilde \rho^\nu)  \nabla \tilde \rho^\nu \right],\;  \big[\sigma (\tilde \rho^\nu) \big]^{-1} \; \left[ \nabla^{\perp} \varphi + {2 D} (\tilde \rho^\nu) \nabla \tilde \rho^\nu \right] \right\rangle  \, dt d\bx \right\}
\end{split}
\label{eq:eqenplus}
\end{equation}
where we recall that $\Omega_L =[0,T] \times [-L,L]^2$. The optimal $\varphi^\nu$ is a solution{\footnote{Two different solutions differ by a function of time which is irrelevant in the variational formula.}} of the PDE
\begin{equation}
\div^{\perp} \left( \sigma^{-1} \big[ \nabla^{\perp} \varphi^\nu + {2 D}(\tilde \rho^\nu)  \nabla \tilde \rho^\nu  \big]\right)=0, \quad \div^\perp=\langle \nabla^{\perp} , \cdot\rangle . 
\end{equation}
The solution $\varphi^\nu$ is in the form of a traveling wave
\begin{equation}
\label{eq:varphinu}
\varphi^\nu (t,\bx)= V^\nu (-\langle \bk,\bx \rangle + v t)
\end{equation}
where $V^\nu$ shall satisfy
\begin{equation}
\begin{split}
&\cfrac{d}{dz} \left[ \left\langle \bk^{\perp} \;  , \;  \big[\sigma(g^\nu)\big]^{-1} \left[  \cfrac{d V^\nu}{dz}  (z) \bk^{\perp} + 2 \cfrac{d g^\nu}{dz} (z) {D} (g^\nu (z))  \bk \right]\right\rangle\right] =0.
\end{split}
\end{equation}
It follows that 
\begin{equation}
\cfrac{ d V^{\nu}}{dz}  (z) = -2\, \cfrac{\langle \bk^{\perp} , \sigma^{-1} {D} \bk \rangle }{\langle \bk^{\perp} , \sigma^{-1} \bk^{\perp} \rangle } \; \cfrac{d g^\nu}{dz} (z) + K
\label{eq:Vprime}
\end{equation}
where $K$ is a constant that has to be optimized. To simplify notation we write ${D},\sigma \ldots$ for ${D} (g^\nu), \sigma(g^\nu) \ldots$. We get then
\begin{equation}
\label{eq:Vprime2}
 \nabla^{\perp} \varphi + {2 D} \nabla \tilde \rho^\nu =  2\,  \cfrac{d g^\nu}{dz} (-\langle \bk, \bx \rangle + v t) \; \left\{ \cfrac{\langle \bk^{\perp} , {\sigma}^{-1} {D} \bk \rangle }{\langle \bk^{\perp} , \sigma^{-1} \bk^{\perp} \rangle } \bk^\perp  -{D} \bk \right\} -K \bk^\perp.
\end{equation}
As shown in Appendix  \ref{sec:app-lim}, optimizing over $K$ yields $K=0$. Therefore 
\begin{equation}
\label{eq:jura79}
\begin{split}
&\nu {H}^{\nu,L} (\tilde \rho^\nu) =\cfrac{1}{\nu L} \; \int_0^{T} \int_{[-L,L]^2} \Big[\cfrac{d g^\nu}{dz} (-\langle \bk, \bx\rangle +vt) \Big]^2 F (g_\nu (-\langle \bk, \bx\rangle +vt ) \; dt d\bx
\end{split}
\end{equation}
where
\begin{equation*}
F(g):=\left\langle \left[ \cfrac{\langle \bk^{\perp} , {\sigma}^{-1} {D} \bk \rangle }{\langle \bk^{\perp} , \sigma^{-1} \bk^{\perp} \rangle } \bk^\perp  -{D} \bk \right] \; , \; \sigma^{-1} \; \left[ \cfrac{\langle \bk^{\perp} , {\sigma}^{-1} {D}  \bk \rangle }{\langle \bk^{\perp} , \sigma^{-1} \bk^{\perp} \rangle } \bk^{\perp} -{D} \bk \right] \right\rangle.
\end{equation*}
The limit of the RHS of this expression as $\nu \to \infty$ and then $L\to \infty$ is postponed to Appendix \ref{sec:app-lim}. We obtain then 
\begin{equation}
\label{eq:jura78}
\lim_{L\to\infty} \tfrac{1}{L} \;  \lim_{\nu \to \infty} {\nu} {H}^{\nu, L} (\tilde \rho^\nu) \; {=}\; 2T \;\int_{\rho^- \wedge \rho^+ }^{\rho^- \vee\rho^+} \,  \cfrac{\langle {\bk} , D(g) \, \bk \rangle}{\langle {\bk}, \sigma(g) \, {\bk} \rangle } \; \Gamma_{\bk}^+ (g,\rho^-,\rho^+) \, dg. 
\end{equation}
This proves the upper bound \eqref{eq:JVgen_movingstep} for an entropy splittable moving step. It is important to remark that the term which contains $\varphi$ in the optimal current \eqref{eq:currentj} does not vanish in general: it gives a non trivial contribution along the shock. However, it does vanish in the special case when $D$ and $\sigma$ are proportional, as is clear from \eqref{eq:Vprime} using that $\sigma^{-1} D$ is a scalar. This is a qualitative difference between the cases when the Einstein relation is satisfied and when it is not.

\subsubsection{Upper bound for a general moving step}
\label{sec:upperbound_general}

We have shown the upper bound \eqref{eq:JVgen_movingstep}, or equivalently \eqref{eq:JVgen_es}, for an entropy splittable moving step. We now argue that 
\eqref{eq:JVgen_movingstep} is valid for a general moving step. Let $\rho(t,\bx)$ be a moving step between $\rho^-$ and $\rho^+$, with direction given by a unit vector $\bk$,  which is not entropy splittable, but can be ``decomposed'' into one totally anti-entropic step, between $\rho^-$ and $\rho_0$, and one entropic step, between $\rho_0$ and $\rho^+$. This means that there exists a unique $\rho_0 \in (\rho^-, \rho^+)$ such that $\Gamma_\bk (\rho^-, \rho^+, \rho_0)=0$ and 
\begin{equation}
\label{eq:aaas}
\Gamma_\bk (g,\rho^-, \rho^+) > 0, \quad g \in (\rho^-, \rho_0); \quad  \Gamma_\bk (g, \rho^-, \rho^+) < 0, \quad g \in (\rho_0, \rho^+).\end{equation}
It implies in particular that
\begin{equation}
\label{eq:split}
\langle (\rho^- -\rho_0) (f(\rho^+)-f(\rho_0))\, -\, (\rho^+ -\rho_0) ( f(\rho^-) -f (\rho_0))\, ,\, \bk  \rangle =0.
\end{equation}

Note that the Rankine-Hugoniot condition together with \eqref{eq:split} impose that the velocity of the steps $(\rho^-,\rho_0)$, $(\rho_0,\rho^+)$, and $(\rho^-,\rho^+)$ are all equal.
As shown in the appendix \ref{sec:appendixB}, it is then relatively straightforward to construct a profile $\rho^{\nu,\delta}$ which approximates the non entropic moving step, and such that
\begin{equation}
\limsup \, \tfrac{1}{2L} \, H^{\nu,L}_{[0,T]} (\rho^{\nu,\delta}) \leq {\bar H} (\rho_1) = {\bar H} (\rho)
\end{equation}
where $\rho_1$ denotes the (entropy splittable) step function between $\rho-$ and $\rho_0$ propagating at velocity $v$ in the direction $\bk$. Here the limsup refers to the ordered limits $\nu \to \infty$, $\delta\to 0$, $L \to \infty$. The equality between $\bar H (\rho)$ and $\bar H (\rho_1)$ is a consequence of (\ref{eq:aaas}).This proves the limsup bound for the non entropy splittable step $\rho$. Clearly, this extends to any step $\rho$ such that the segment $[\rho^-,\rho^+]$ can be decomposed into entropic and totally anti-entropic intervals.

\subsubsection{Lower bound for a moving step}
\label{sec:lbmsf}
In order to get a lower bound for $\bar H$ we have to compute 
\[
\lim_{L \to \infty} \tfrac{1}{2L} \,\sup_{{\mc V} \in {\bf {\hat V}_{\rho}}^L} {\mc P}_{\mc V} (\rho)
\]
for $\rho$ a moving step function, with direction $\bk$, $\|\bk\|=1$, and velocity $v$, i.e.
\begin{equation*}
\rho(t,\bx)=
\begin{cases}
& \rho^- \quad \mbox{if}~\langle \bk,\bx\rangle \leq vt  \nonumber \\
& \rho^+ \quad \mbox{if}~\langle \bk,\bx\rangle > vt. \nonumber
\end{cases}
\end{equation*}
Here $ {\bf {\hat V}_{\rho}}^L$ is the set of entropy samplers ${\mc V} \in {\bf {\hat V}_{\rho}}$ vanishing outside of $\Omega_L=[0,T]\times [-L, L]^2$ (see Section \ref{sec:lowerbounds} for a definition of ${\hat V}_{\rho}$). Exploiting
that for any $(t,\bx)$ in the discontinuity set $n(t,\bx)=\bk$ and that ${\mc V} \in {\bf {\hat V}_{\rho}}^L$ we have 
$${\mc V}'' (g,t, \bx) \le 2 \cfrac{\langle D(g) \bk, \bk\rangle}{\langle \sigma (g) \bk, \bk \rangle}\; {\bf 1}_{(t,\bx) \in \Omega_L}.$$
It follows that
\begin{equation*}
\begin{split}
{\mc P}_{\mc V} (\rho) & \le \int \;  \int_{J_{\rho} \cap \Omega_L} d\gamma_{J_\rho}\;  {\mc V}'' (g,t,\bx) \{ \mu_\rho^+ (g, dt, d\bx) -\mu_\rho^- (g, dt,d\bx) \} \, dg \\
&\le \int  \;  \int_{J_{\rho} \cap \Omega_L} d\gamma_{J_\rho} \;  {\mc V}'' (g,t,\bx) \mu_\rho^+ (g, dt, d\bx) \, dg\\
&\le 2 \int \;  \cfrac{\langle D(g) \bk, \bk\rangle}{\langle \sigma (g) \bk, \bk \rangle} \;  \int_{J_{\rho} \cap \Omega_L} d\gamma_{J_\rho}\;  \mu^+ (g, dt, d\bx) \, dg\\
&=  2 \int \;  \cfrac{\langle D(g) \bk, \bk\rangle}{\langle \sigma (g) \bk, \bk \rangle} \; \mu_\rho^+ (g, \Omega_L) \, dg\\
&=2 \int \;  \cfrac{\langle D(g) \bk, \bk\rangle}{\langle \sigma (g) \bk, \bk \rangle} \; \mu_\rho^+ (g, E_\rho^+ (g) \cap \Omega_L )\, dg.
\end{split}
\end{equation*}
The inequalities above become equalities with the choice 
$${\mc V}^{\prime\prime} (g, t, \bx)=2  \cfrac{\langle D(g) \bk, \bk\rangle}{\langle \sigma (g) \bk, \bk \rangle} {\bf 1}_{ (t,\bx) \in E_\rho^+ (g) \cap \Omega_L},$$ which is possible in ${\bf {\hat V}_{\rho}}^L$; notice that such a choice is not necessarily admissible in ${\bf \hat{V}}$, hence the lower bound \eqref{eq:liminf} may not be strong enough. Therefore we have that
\begin{equation}
\sup_{{\mc V} \in {\bf {\hat V}_{\rho}}^L} {\mc P}_{\mc V} (\rho)=2 \int \;   \cfrac{\langle D(g) \bk, \bk\rangle}{\langle \sigma (g) \bk, \bk \rangle} \; \mu_\rho^+ (g, E_\rho^+ (g) \cap \Omega_L) \, dg.
\end{equation}
It follows that for a moving step function propagating in the direction $\bk$ we have
\begin{equation}
\begin{split}
\liminf_{\nu \to \infty} \, \cfrac{1}{2L} \, H_{[0,T]}^{\nu, L} (\rho)& \ge \, \cfrac{1}{L} \;\int_{J_\rho \cap \Omega_L} \,  \theta^+ (g, \rho^-, \rho^+)\, d\gamma_{J_\rho} \, \int \, \cfrac{\langle D(g) \bk, \bk\rangle}{\langle \sigma (g) \bk, \bk \rangle} \, dg\\
&=\cfrac{| J_\rho \cap \Omega_L|}{L} \, \int_{\rho^- \wedge \rho^+}^{\rho^- \vee \rho^+}  \cfrac{\langle D(g) \bk, \bk\rangle}{\langle \sigma (g) \bk, \bk \rangle} \,  \Gamma_{\bk}^+ (g, \rho^-, \rho^+)\, dg
\end{split}
\end{equation}
where $|J_\rho \cap \Omega_L|=2T L$ is the area of the discontinuity set of the weak solution $\rho$ in $\Omega_L$. Taking the limit $L\to \infty$ this proves the formula for the lower bound of $\bar H (\rho)$.

Finally, this result together with that of Section \ref{sec:upperbound_general} shows the announced formula for a general moving step.

\subsection{The upper bound in \eqref{eq:conj} and the additive principle in two dimensions}
\label{sec:additivity}
Now that we have proved the upper bound for all moving step functions, we imagine we can approximate a general weak solution $\rho$ as a superposition of local moving steps. Using the additive principle of Section \ref{sec:additive_principle}, we can deduce the upper bound in \eqref{eq:conj} for a general weak solution.

Nevertheless, there is a potential difficulty here. The key point is that the optimal current \eqref{eq:currentj} associated to a step has a trivial part (which is the flux $f$ in the large $\nu$ limit), plus a non trivial part which essentially vanishes except close to the step. Thus, it is possible that there is no interference between different discontinuities, and that the costs simply add up. This picture seems correct for two steps propagating in the same direction at the same speed (Section \ref{sec:upperbound_general}); nevertheless, since the non trivial part of the optimal current does not vanish along the step (see \eqref{eq:varphinu} and \eqref{eq:Vprime}), it is a bit less clear for more general discontinuities, where everything (shock direction, height, velocity...) varies continuously.

\section*{Acknowledgements}
We acknowledge very useful discussions with C. Bahadoran, T. Bodineau, M. Mariani and C. Nardini.
This work has been supported by the Brazilian-French Network in Mathematics and the project EDNHS ANR-14-CE25-0011 of the French National Research Agency (ANR) and the project LSD ANR-15-CE40-0020-01 LSD of the French National Research Agency (ANR). This research was supported in part by the International Centre for Theoretical Sciences (ICTS) during a visit for participating in the program Non-equilibrium statistical physics (Code: ICTS/Prog-NESP/2015/10).

\appendix 

\section{Proof of (\ref{eq:thetachaise}) }
\label{sec:appendixA}
The relations (\ref{eq:thetachaise},\ref{eq:muvu2}) give
\begin{equation}
\label{eq:thetachaise}
\begin{split}
&\theta (p,\rho^-,\rho^+) = {\bf 1}_{ \inf(\rho^-,\rho^+) \le p \le \sup(\rho^-, \rho^+)} \\
& \quad \quad \quad \times {\rm{sgn}} (\rho^+ -\rho^-) \; \left\{ (\rho^- -p) {\bf n}^t +  \langle f(\rho^-) -f(p)\, , \,  {\bf n}^\bx \rangle \right\}\\
&= {\bf 1}_{ \inf(\rho^-,\rho^+) \le p \le \sup(\rho^-, \rho^+)}\\
&\times \cfrac{1}{|\rho^+ -\rho^-|} \; \left\langle f(\rho^-) (\rho^+ -p) + f(\rho^+) (p-\rho^-) -f(p) (\rho^+ -\rho^-) \, , \, {\bf n}^\bx \right\rangle\\
&={\bf 1}_{ \inf(\rho^-,\rho^+) \le p \le \sup(\rho^-, \rho^+)}  \; \Gamma_{\tiny{\tfrac{{\bf n}^{\bx}}{\| {\bf n}^\bx \|}}} (p, \rho^-,\rho^+) \; \| {\bf n}^\bx\|
\end{split}
\end{equation}

where $\Gamma_{\bk} (p, \rho^-,\rho^+)$, $\|\bk \|=1$ is defined by  (\ref{eq:thetachaisegamma}).


\section{Approximating a non entropy splittable moving step}
\label{sec:appendixB}

The goal is here to provide an approximating profile for a non entropy splittable moving step as in Section \ref{sec:upperbound_general}, and compute its cost. We introduce the approximating profiles for each one of the smaller steps $\rho_i^\nu$, $i=1,2$, as traveling waves $\rho^\nu_i (t,\bx) =g_i^\nu( \langle k,\bx\rangle -vt)$, solutions of the equations
\begin{eqnarray*}
\partial_t \rho_1^\nu + \div f(\rho_1^\nu)&=& -\frac{1}{\nu}\nabla (D(\rho_1^\nu)\nabla \rho_1^\nu)\\
\partial_t \rho_2^\nu + \div f(\rho_2^\nu)&=& \frac{1}{\nu}\nabla (D(\rho_2^\nu)\nabla \rho_2^\nu)
\end{eqnarray*}
with boundary conditions 
$$\lim_{z \to -\infty} g_1^\nu (z)=\rho^-, \quad  \lim_{z \to +\infty} g_1^\nu (z)=\rho_0,$$
$$\lim_{z \to -\infty} g_2^\nu (z)=\rho_0, \quad  \lim_{z \to +\infty} g_2^\nu (z)=\rho^+.$$
We choose the $g_i^\nu$ such that the strong gradient of $g_i^\nu(z)$ is around $z=0$. For $|z| \, \gg \, 1/\nu$, the $g_i^\nu$ are exponentially close to their asymptotic value. To be more precise, we have for $z>0$ and $\nu z$ which tends to infinity:
\[
g_i^\nu( \nu z) -g_i^\nu(+ \infty) \sim A e^{-\lambda \, \nu z  }
\]
for some positive constant $\lambda$ of order $1$. A similar exponential estimate holds for negative $z$.

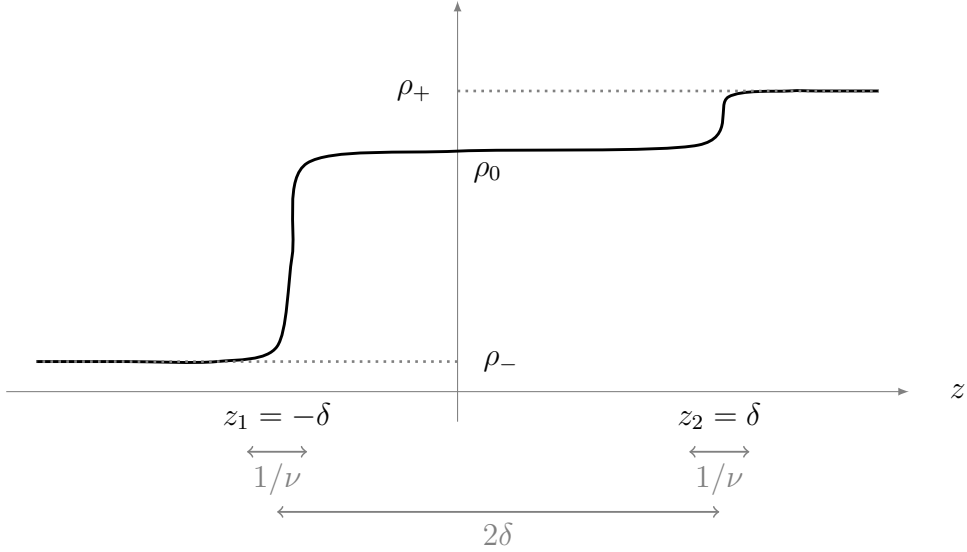
\begin{figure}
\begin{center}
\begin{tikzpicture}[scale=0.4]
\draw[->,>=latex, Gray] (-15,0) -- (15,0);
\draw[->,>=latex, Gray] (0,-1) -- (0,13);
\draw [line width=1.1pt, black,smooth] plot coordinates {(-14,1) (-11,1) (-8,1) (-6,1.5) (-5.5, 4.5) (-5,7.6) (0,8) (8,8.2) (9,9.8) (11,10) (12,10) (14,10)};
\node[right, black] at (16,0) {$z$} ;
\node[below, black] at (-6,0) {$z_1=-\delta$} ;
\node[below, black] at (8.7,0) {$z_2=\delta$} ;
\node[right, below, black] at (1,8) {$\rho_0$} ;
\node[right, black] at (0.5,1) {$\rho_-$} ;
\node[left, black] at (-0.5,10) {$\rho_+$} ;
\draw[-,dotted, line width=1 pt, Gray] (0,10)--(14,10);
\draw[-,dotted, line width=1 pt, Gray] (0,1)--(-14,1);
\draw[<->, line width=0.7pt, Gray] (-7,-2)--(-5,-2) node[midway, below, sloped] {$1/\nu$};
\draw[<->, line width=0.7pt, Gray] (7.7,-2)--(9.7,-2) node[midway, below, sloped] {$1/\nu$};
\draw[<->, line width=0.7pt, Gray] (-6,-4)--(8.7,-4) node[midway, below, sloped] {$2\delta$};
\end{tikzpicture}
\end{center}
\caption{ The approximating profile $g^{\nu,\delta} (z)$.}
\label{fig:3}
\end{figure}

We also introduce a new small parameter $\delta$, and the shifted approximating profiles:
\[
\rho_i^{\nu,\delta} (t,\bx)  = \rho_i^\nu(t+\delta/v,\bx).
\]
Let $\chi_\delta(z)$ be a step function increasing from $0$ to $1$ around $z=0$, regularized at scale $\delta$, with compact support in $]-\delta/2,\delta/2[$; we take $\chi_\delta$ such that
$\chi_\delta(-z)+\chi_\delta(z)=1$. We write now an approximation for the whole step:
\begin{equation*}
\begin{split}
\rho^{\nu,\delta}(t, \bx) &=g^{\nu,\delta} (\langle \bk, \bx \rangle -vt)\\
&= \chi_\delta(- \langle \bk,\bx\rangle +vt)\, \rho_1^{\nu,-\delta} (t,\bx)+\chi_\delta(\langle \bk,\bx\rangle-vt)\, \rho_2^{\nu,\delta}(t,\bx).
\end{split}
\end{equation*}
If $\delta$ is much larger than $1/\nu$, $\rho^{\nu,\delta}$ is thus a "double step", see Figure \ref{fig:3}, which varies at scale $1/\nu$ around $z_1=\langle \bk,\bx\rangle-vt=-\delta$, and around
$z_2=\langle \bk,\bx\rangle-vt=\delta$. In the following, we shall always bear in mind this ordering: $1/\nu\,  \ll \, \delta \, \ll \, 1$.

Notice that $\rho^{\nu,\delta} (t, \bx)=\rho_1^{\nu,-\delta} (t,\bx)$ (resp. $=\rho_2^{\nu,\delta} (t,\bx)$) when $z=\langle \bk,\bx\rangle-vt\leq -\delta/2$ (resp. $z\geq \delta/2$). 
For $-\delta/2\leq z\leq \delta/2$, $\rho_1^{\nu,-\delta} (t,\bx)$ and $\rho_2^{\nu,\delta}(t,\bx)$ are both very close to $\rho_0$, that is exponentially in $\nu \delta$: these are the tails of the traveling waves profiles. We have that
\[
\partial_t \rho^{\nu,\delta} = -\partial_t z \, \chi'_\delta(-z)\rho_1^{\nu,-\delta} + \partial_t z \, \chi'_\delta(z)\rho_2^{\nu,\delta} 
+\chi_\delta(-z)\, \partial_t \rho_1^{\nu,-\delta}+\chi_\delta(z) \, \partial_t \rho_2^{\nu,\delta}.
\]
Making use of the fact that $\chi'_\delta(-z)-\chi'_\delta(z) =0$, we see that between the two sub-steps, that is for $-\delta/2\leq z\leq \delta/2$, 
$\partial_t \rho^{\nu,\delta}$ is exponentially small in $\nu \delta$. For $z\leq -\delta/2$ (resp. $ z\geq \delta/2$), $\partial_t \rho^{\nu,\delta}$ reduces to the contribution coming from $\rho_1^{\nu,-\delta}$ (resp. $\rho_2^{\nu,\delta}$). To summarize
\begin{equation}
\partial_t \rho^{\nu,\delta} =
\begin{cases}
 -\div [ f(\rho_1^{\nu,-\delta}) + D\nabla \rho_1^{\nu,-\delta}]~{\rm for}~  z\leq -\delta/2, \\
{\rm exponentially~small~in~} \nu\delta ~{\rm for}~  -\delta/2\leq z\leq \delta/2, \\
- \div [ f(\rho_2^{\nu,\delta}) - D\nabla \rho_2^{\nu,\delta}]~{\rm for}~  z\geq \delta/2.  
\end{cases}
 \label{eq:dtrhonu}
\end{equation}
Similarly, we have that
\begin{equation}
\nabla \rho^{\nu,\delta} =
\begin{cases}
\nabla \rho_1^{\nu,-\delta}~{\rm for}~  z\leq -\delta/2, \\
{\rm exponentially~small~in~} \nu\delta~{\rm for}~  -\delta/2\leq z\leq \delta/2, \\
\nabla \rho_2^{\nu,\delta}~{\rm for}~  z\geq \delta/2.
\end{cases} 
\label{eq:nablarhonu}
\end{equation}
We can now estimate $H^{\nu,L}_{[0,T]} (\rho^{\nu,\delta})$, starting from the expression
\begin{equation}
H^{\nu,L}_{[0,T]} (\rho^{\nu,\delta}) = \frac{1}{2} \inf_{j} \int_{\Omega_L} \Big\langle j- j_\nu (\rho^{\nu,\delta})\,,\,\sigma^{-1} (\rho^{\nu,\delta})\,\, (j-j_\nu (\rho^{\nu,\delta})) \Big\rangle \, dt d\bx
\label{eq:Hnudoublestep}
\end{equation}
with the constraint $\partial_t\rho^{\nu,\delta}+\div j=0$. 
We take $j=f(\rho^{\nu,\delta})+b$ and want to construct the optimal $b$. Using \eqref{eq:dtrhonu}, 
the constraint becomes:
\begin{equation*}
\div b=
\begin{cases}
\div (D(\rho_1^{\nu,-\delta})\nabla \rho_1^{\nu,-\delta}) ~{\rm for}~  z\leq -\delta/2, \nonumber \\
{\rm exponentially~small~in~} \nu\delta   ~{\rm for}~  -\delta/2\leq z\leq \delta/2 \nonumber \\
- \div (D(\nabla \rho_2^{\nu,\delta})\nabla \rho_2^{\nu,\delta})  ~{\rm for}~  z\geq \delta/2. \nonumber
\end{cases}
\end{equation*}
The idea is now to choose
\begin{equation*} 
b^{\nu,\delta}=
\begin{cases}
b_1^{\nu,\delta}=D(\rho_1^{\nu,-\delta})\nabla \rho_1^{\nu,-\delta} +\frac{1}{\nu}\nabla^\perp \varphi^{\nu,-\delta} ~{\rm for}~  z\leq -\delta/2, \nonumber \\
{\rm exponentially~small~in~} \nu\delta   ~{\rm for}~  -\delta/2\leq z\leq \delta/2, \nonumber \\
b_2^{\nu,\delta}=-D(\nabla \rho_2^{\nu,\delta})\nabla \rho_2^{\nu,\delta} ~{\rm for}~  z\geq \delta/2, \nonumber
\end{cases}
\end{equation*} 
where $\nabla^\perp \varphi^{\nu,-\delta}$ is the optimal non trivial contribution to the current computed in Section \ref{UB_entrop_split} for the totally anti-entropic moving step, shifted by $-\delta$. The vector field $b_1^{\nu,\delta}$ contains the gradient-like part $D(\rho_1^{\nu,-\delta})\nabla \rho_1^{\nu,-\delta}$, plus a rotational part. Notice that both $b_1^{\nu,\delta}(z)$ and $b_2^{\nu,\delta}(z)$ vanish exponentially in the intermediate region $-\delta/2\leq z\leq \delta/2$; hence $b^{\nu,\delta}$ as constructed above can be chosen to be smooth.
Inserting $b^{\nu,\delta}$ in \eqref{eq:Hnudoublestep}, the $b_1^{\nu,\delta}$ part contributes the cost of the totally anti-entropic step (up to small terms), and $b_2^{\nu,\delta}=- D(\rho_2^{\nu, \delta})\nabla \rho_2^{\nu,\delta}$ contributes only small terms (that is essentially the cost of the entropic step).
Then, by denoting $\rho_1$ the (entropy splittable) step function between $\rho^-$ and $\rho_0$ propagating at velocity $v$ in the direction $\bk$, we get
\begin{equation*}
\limsup \tfrac{\nu}{2L} H^{\nu,L}_{[0,T]} (\rho^{\nu,\delta}) \leq {\bar H}(\rho_1) = {\bar H} (\rho)
\end{equation*}
where the limsup refers to the ordered limits $\nu \to \infty$, $\delta \to 0$ and then $L\to \infty$. This proves the limsup bound for \textcolor{blue}{this simple} non entropy splittable step. 

\section{Proof of (\ref{eq:jura78}) }
\label{sec:app-lim}
We first show that the choice $K=0$ is optimal.
Introducing the notation ${D} \bk =\alpha \bk^\perp +\beta \bk$, we notice that
\begin{equation}
\label{eq:jolie}
\cfrac{\langle \bk^{\perp} , {\sigma}^{-1} {D} \bk \rangle }{\langle \bk^{\perp} , \sigma^{-1} \bk^{\perp} \rangle } \bk^\perp  -{D} \bk = \alpha 
\cfrac{\langle \bk^{\perp} , {\sigma}^{-1} \bk \rangle }{\langle \bk^{\perp} , \sigma^{-1} \bk^{\perp} \rangle }\bk^\perp -\alpha \bk. 
\end{equation}
Hence
\begin{equation}
\label{eq:grosprodscal}
\Bigg \langle \left\{ \cfrac{\langle \bk^{\perp} , {\sigma}^{-1} {D} \bk \rangle }{\langle \bk^{\perp} , \sigma^{-1} \bk^{\perp} \rangle } \bk^\perp  -{D} \bk \right\} , \sigma^{-1}\bk^\perp\Bigg \rangle = \alpha \frac{\langle \bk^{\perp} , {\sigma}^{-1} \bk \rangle}{\langle \bk^{\perp} , {\sigma}^{-1} \bk^\perp \rangle} \langle \bk^{\perp} , {\sigma}^{-1} \bk^\perp \rangle-\alpha \langle \bk , {\sigma}^{-1} \bk^\perp \rangle =0.
\end{equation}
When \eqref{eq:Vprime2} is introduced into \eqref{eq:eqenplus}, the term which is linear in $K$ contains precisely the scalar product \eqref{eq:grosprodscal}; thus it vanishes.
The term which is quadratic in $K$ is $\langle \bk^\perp,\sigma^{-1} \bk^\perp\rangle$, a strictly positive quantity. Hence it is minimized for $K=0$, as announced. 

Now, the integral of (\ref{eq:jura79}) is equal to
\begin{equation*}
\begin{split}
& \cfrac{1}{\nu L} \int_0^T \int_{[-L,L]^2} {\Big[\cfrac{d g^\nu}{dz} (-\langle \bk, \bx\rangle +vt)\Big]^2}  F(g_\nu (-\langle \bk, \bx\rangle +vt) \; dt d\bx\\
&= \cfrac{1}{v \nu L} \int_{[-L,L]^2} d\bx \; \int_{-\langle \bk, \bx\rangle }^{-\langle \bk, \bx\rangle +vT } {\Big[\cfrac{d g^\nu}{dz} (z)\Big]^2}\;  F[g_\nu (z)] \; dz
\end{split}
\end{equation*}

By multiplying (\ref{eq:5hmat}) by $F(g_\nu) \tfrac{d  g^\nu}{dz}$ and integrating between ${-\langle \bk, \bx\rangle }$ and ${-\langle \bk, \bx\rangle +vT }$ we conclude that
\begin{equation*}
\begin{split}
&\cfrac{1}{\nu v} \int_{-\langle \bk, \bx\rangle }^{-\langle \bk, \bx\rangle +vT } {\Big[\cfrac{d g^\nu}{dz} (z)\Big]^2}\; F[g_\nu (z)] \; dz \\
& = \cfrac{1} {v} \; \left\langle \int_{g^\nu(-\langle \bk, \bx\rangle) }^{g^\nu(-\langle \bk, \bx\rangle +vT)} ( f(g) - v g \bk + C_{\infty} \bk) \, \cfrac{F(g)}{\langle \bk, {D} \bk \rangle} \, dg \; , \; \bk \right\rangle.
\end{split}
\end{equation*}
This converges as $\nu \to \infty$ to
 \begin{equation*}
\begin{split}
\cfrac{1} {v} \; \left\langle \int_{g^\infty(-\langle \bk, \bx\rangle) }^{g^\infty(-\langle \bk, \bx\rangle +vT)} ( f(g) - v g \bk + C_{\infty} \bk) \, \cfrac{F(g)}{\langle \bk, {D}(g) \bk \rangle} \, dg \; , \; \bk \right\rangle.
\end{split}
\end{equation*}
Observe this is non zero if and only if the shock $z_s$ is in the interval $[\inf \{-\langle \bk, \bx\rangle, -\langle \bk, \bx\rangle +vT\}\; ; \; \sup\{-\langle \bk, \bx\rangle, -\langle \bk, \bx\rangle +vT \}]$. Therefore we get that
\begin{equation}
\begin{split}
\lim_{\nu \to \infty}  \nu{H}^{\nu, L} (\tilde \rho^\nu)&= \left| \left\langle  \int_{\rho^- }^{\rho^+ }( f(g) - v g \bk + C_{\infty} \bk)\, \cfrac{F(g)} {\langle \bk, {D}(g) \bk \rangle} \, dg \; , \; \bk \right\rangle \right|\\
&\times  \cfrac{1}{|v|L} \int_{[-L,L]^2} d\bx \; {\bf 1}_{\left\{ z_s \in \inf \{-\langle \bk, \bx\rangle, -\langle \bk, \bx\rangle +vT\}\; ; \; \sup\{-\langle \bk, \bx\rangle, -\langle \bk, \bx\rangle +vT \}] \right\}}.
\end{split}
\end{equation}
To conclude we observe that
\begin{equation*}
\begin{split}
&\cfrac{F(g)} {\langle \bk, { D}(g) \bk \rangle}\\
&=\cfrac{1} {\langle \bk, {D}\bk \rangle} \left\langle \left[ \cfrac{\langle \bk^{\perp} , {\sigma}^{-1} {D} \bk \rangle }{\langle \bk^{\perp} , \sigma^{-1} \bk^{\perp} \rangle } \bk^\perp  -{D} \bk \right] \; , \; \sigma^{-1} \; \left[ \cfrac{\langle \bk^{\perp} , {\sigma}^{-1} {D}  \bk \rangle }{\langle \bk^{\perp} , \sigma^{-1} \bk^{\perp} \rangle } \bk^{\perp} -{D} \bk \right] \right\rangle.
\end{split}
\end{equation*}
Making use of \eqref{eq:jolie} and of $\alpha = \langle \bk \, , \, { D} \bk \rangle$, we have
\begin{equation*}
\begin{split}
\cfrac{F(g)} {\langle \bk, { D}(g) \bk \rangle}&= \cfrac{\langle \bk \, , \, D \bk \rangle}{\langle \bk^{\perp} , {\sigma}^{-1} \bk^\perp \rangle} \; \left\{ \langle \bk , {\sigma}^{-1} \bk \rangle \langle \bk^{\perp} , {\sigma}^{-1} \bk^{\perp}\rangle - \langle \bk^{\perp} , {\sigma}^{-1} \bk \rangle^2\right\}\\
&= \cfrac{\langle \bk \, , \, D \bk \rangle}{\langle \bk^{\perp} , {\sigma}^{-1} \bk^\perp \rangle} \; \left\{ \langle \bk , {\sigma}^{-1} \bk \rangle \langle \bk^{\perp} , {\sigma}^{-1} \bk^{\perp}\rangle - \langle \bk^{\perp} , {\sigma}^{-1} \bk \rangle^2\right\} \\
&= \;  \cfrac{\langle \bk \, , \, D \bk \rangle}{\langle \bk^{\perp} , {\sigma}^{-1} \bk^\perp \rangle} \; {\rm{det}} (\sigma^{-1}) =  \;  \cfrac{\langle \bk \, , \, D \bk \rangle}{\langle \bk , {\sigma} \bk \rangle}.
\end{split}
\end{equation*}
We conclude that
\begin{equation*}
\begin{split}
\lim_{\nu \to \infty} \nu {H}^{\nu, L} (\tilde \rho_\nu)&= \, \left| \left\langle  \int_{\rho^- }^{\rho^+ } \;  \cfrac{\langle \bk \, , \, D \bk \rangle}{\langle \bk , {\sigma} \bk \rangle} \, ( f(g) - v g \bk + C_{\infty} \bk)\, dg \; , \; \bk \right\rangle \right|\\
&\times \cfrac{1}{ |v| L} \int_{[-L,L]^2} d\bx \; {\bf 1}_{\left\{ z_s \in \inf \{-\langle \bk, \bx\rangle, -\langle \bk, \bx\rangle +vT\}\; ; \; \sup\{-\langle \bk, \bx\rangle, -\langle \bk, \bx\rangle +vT \}] \right\}}
\end{split}
\end{equation*}
We then use (\ref{eq:cinfty}) to get that
\begin{equation}
\begin{split}
&\left\langle \int_{\rho^- }^{\rho^+ } \;  \cfrac{\langle \bk \, , \, D \bk \rangle}{\langle \bk , {\sigma} \bk \rangle} \, ( f(g) - v g \bk + C_{\infty} \bk)\,  dg \; , \; \bk \right\rangle\\
&=\int_{\rho^- }^{\rho^+ }   \cfrac{\langle \bk \, , \, D \bk \rangle}{\langle \bk , {\sigma} \bk \rangle}  \,  \cfrac{\left\langle  (\rho^+ -\rho^-) f(g) +(g-\rho^+) f(\rho^-) +(\rho^- -g) f(\rho^+) \; , \bk \right\rangle}{\rho^+ -\rho^-}\, dg\\
&=\int_{\rho^- }^{\rho^+ }   \cfrac{\langle \bk \, , \, D \bk \rangle}{\langle \bk , {\sigma} \bk \rangle}  \, \Gamma_{\bk} (g, \rho^-,\rho^+) \, dg.
\end{split} \label{eq:jura78_proved}
\end{equation}
Observe also that
\begin{equation*}
\lim_{L \to \infty } \cfrac{1}{2|v|L} \int_{[-L,L]^2} d\bx \; {\bf 1}_{\left\{ z_s \in \inf \{-\langle \bk, \bx\rangle, -\langle \bk, \bx\rangle +vT\}\; ; \; \sup\{-\langle \bk, \bx\rangle, -\langle \bk, \bx\rangle +vT \}] \right\}} =T.
\end{equation*}
Finally, since we are considering a totally anti entropic moving step, $\Gamma_{\bk}$ coincides with $\Gamma_{\bk}^+$, and \eqref{eq:jura78_proved} coincides with 
\eqref{eq:jura78}, which we wanted to prove.

\section{Proof of (\ref{eq:varfl}) }
\label{sec:app-vf}
To get (\ref{eq:varfl}) observe that the supremum appearing there is realized for $\hat \varphi$ solution to   
$$ \partial_t u + \div j_\nu (u) + \cfrac{1}{\nu} \, \div ( \sigma \nabla \hat \varphi ) =0$$
with zero boundary conditions at infinity. The RHS of (\ref{eq:varfl}) is then equal to 
\begin{equation}
\label{eq:123}
\cfrac{1}{2\nu} \, \int  \langle \nabla \hat \varphi,\sigma \nabla \hat \varphi \rangle \, dt d\bx.
\end{equation}
Observe also that $\hat j = j_\nu (u) + \cfrac{1}{\nu} \, \sigma \nabla \hat \varphi$ satisfies $\partial_{t} u + \div \hat j =0$. On the other hand we have that 
\begin{equation*}
\begin{split}
{\nu}H^\nu (u)&= \inf_{j~s.t.~\partial_t u +\div j=0}  \cfrac{\nu}{2}  \; \int  \left \langle \big[ j - j_\nu (u) \big] \, , \,  \sigma^{-1} (u) \big[ j -j_\nu (u) \big] \right \rangle  dt d\bx\\
&=\cfrac{\nu}{2} \inf_\psi   \; \int  \left \langle \big[ \nabla^{\perp} \psi + \tfrac{1}{\nu}  \sigma \nabla \hat \varphi  \big] \, , \,  \sigma^{-1} (u) \big[ \nabla^{\perp} \psi + \tfrac{1}{\nu}  \sigma \nabla \hat \varphi \big] \right \rangle  dt d\bx\\
\end{split}
\end{equation*}
since any divergence free vector field is a rotational. Since 
$$\int \langle  \nabla^{\perp} \psi\, , \, \nabla \varphi \rangle \, dt d\bx =0$$
we deduce that the previous infimum is realized for constant $\psi$ and equal to (\ref{eq:123}). This proves the claim.

\end{document}